\documentclass[aps,prb,10pt,amsmath,twocolumn,amssymb,superscriptaddress,showpacs]{revtex4-1} 
\usepackage{bm}
\usepackage[]{graphicx}
\usepackage[tight]{units}
\usepackage{color}
\usepackage{ulem}
\usepackage{marvosym}
\usepackage[colorlinks=true,citecolor=blue,linkcolor= blue,linkcolor=blue,urlcolor=blue]{hyperref}

\definecolor{lred}{rgb}{.55,.0,.0}

\newcommand{\BiTe}{$\text{Bi}_2\text{Te}_3$~}

\newcommand{\BiSe}{$\text{Bi}_2\text{Se}_3$~}
\newcommand{\SbTe}{$\text{Sb}_2\text{Te}_3$~}

\newcommand{\BiTex}{$\text{Bi}_2\text{Te}_3$}

\newcommand{\SbTex}{$\text{Sb}_2\text{Te}_3$}
\newcommand{\BiSbTex}{$\text{Bi}_{(2-x)}\text{Sb}_{(x)}\text{Te}_3$}

\newcommand{\f}[1]{Fig.~\ref{fig:#1}}
\newcommand{\fs}[1]{Figs.~\ref{fig:#1}}

\begin{document}

\title[]{Signature of the topological surface state in the thermoelectric properties of $\text{Bi}_2\text{Te}_3$}
\author{F. Rittweger}
\affiliation{Max-Planck-Institut f\"{u}r Mikrostrukturphysik, Weinberg 2, DE-06120 Halle, Germany}
\author{N. F. Hinsche}
\email{nicki.hinsche@physik.uni-halle.de}
\affiliation{Institut f\"{u}r Physik, Martin-Luther-Universit\"{a}t Halle-Wittenberg, DE-06099 Halle, Germany}
\author{P. Zahn}
\affiliation{Helmholtz-Zentrum Dresden-Rossendorf, P.O.Box 51 01 19, DE-01314 Dresden, Germany}
\author{I. Mertig}
\affiliation{Max-Planck-Institut f\"{u}r Mikrostrukturphysik, Weinberg 2, DE-06120 Halle, Germany}
\affiliation{Institut f\"{u}r Physik, Martin-Luther-Universit\"{a}t Halle-Wittenberg, DE-06099 Halle, Germany}

\date{\today}

\begin{abstract}
We present \textit{ab initio} electronic structure calculations based on density functional theory for the thermoelectric
properties of \BiTe films. Conductivity and thermopower are computed in the diffusive limit of transport based on the 
Boltzmann equation. Bulk and surface contribution to the transport coefficients are separated by a special projection
technique. As a result we show clear signatures of the topological surface state in the thermoelectric properties.
\end{abstract}


\maketitle

Recent studies in condensed-matter physics showed that \BiTex, 
which is one of the most studied and efficient thermoelectric 
materials \cite{Bottner:2006p2812,Snyder:2008}, belongs to 
the group of Z2 
topological insulators \cite{Fu:2007p15433,Zhang:2009,Hasan:2010p15275}. 
Clearly the link between an efficient thermoelectric material and the topological 
character is the spin-orbit induced inverted band gap. 
While the small size of the band gap favours room-temperature
thermoelectrics \cite{Mahan:1989}, the inversion itself is 
often triggered by heavy atoms, leading to low lattice thermal 
conductivity enhancing the figure of merit. 
In addition, the existence of topological surface states opens the 
opportunity to increase the performance of 
thermoelectric devices \cite{Hor:2009p15893,Ghaemi:2010,Takahashi:2012}.

While many experiments and calculations were performed 
investigating the robustness and the spin texture of the gapless 
surface state in \BiTex \cite{Hasan:2010p15275,Henk:2012p16289,Ando2013}, the precise identification 
of the surface states contribution to the various transport coefficients 
is still an open question \cite{Qu:2010,Taskin:2011p15963}.
 
We present \textit{ab initio} calculations of the thermoelectric transport properties of a \BiTe film. The 
transport properties of the \BiTe film are calculated in the diffusive limit of transport by means of the 
semiclassical Boltzmann equation in relaxation time approximation (RTA) \cite{Mertig:1999p12776,Hinsche:2011p15276,Thonhauser:2003p14996,Huang:2008p559,Park:2010p11006,Hinsche:2012p16003,Hinsche:2011p15707}. 
Within this approximation we assume that the attached metallic leads basically preserve the surface band structure.

The electronic structure of the \BiTe surface was obtained by first principles density functional theory calculations (DFT), 
as implemented in the fully relativistic screened Korringa-Kohn-Rostoker Green function method (KKR) \cite{Gradhand:2009p7460}. 
Exchange and correlation effects were accurately accounted for by the local density approximation (LDA) parametrized by Vosco, Wilk, and Nusair \cite{Vosko1980}.
The atomistic structure was simulated by a slab configuration of 20 atomic layers \BiTex, i.e. 
four quintuple layers (QL), separated by a vacuum spacer of sufficient thickness to \FR{separate} the films. The experimental 
in-plane lattice parameter ${a^{\text{hex}}_{\text{BiTe}}}=4.384${\AA } and relaxed atomic positions \cite{Landolt} were used. 
\begin{figure*}[t]
\centering
\includegraphics[width=0.9\textwidth]{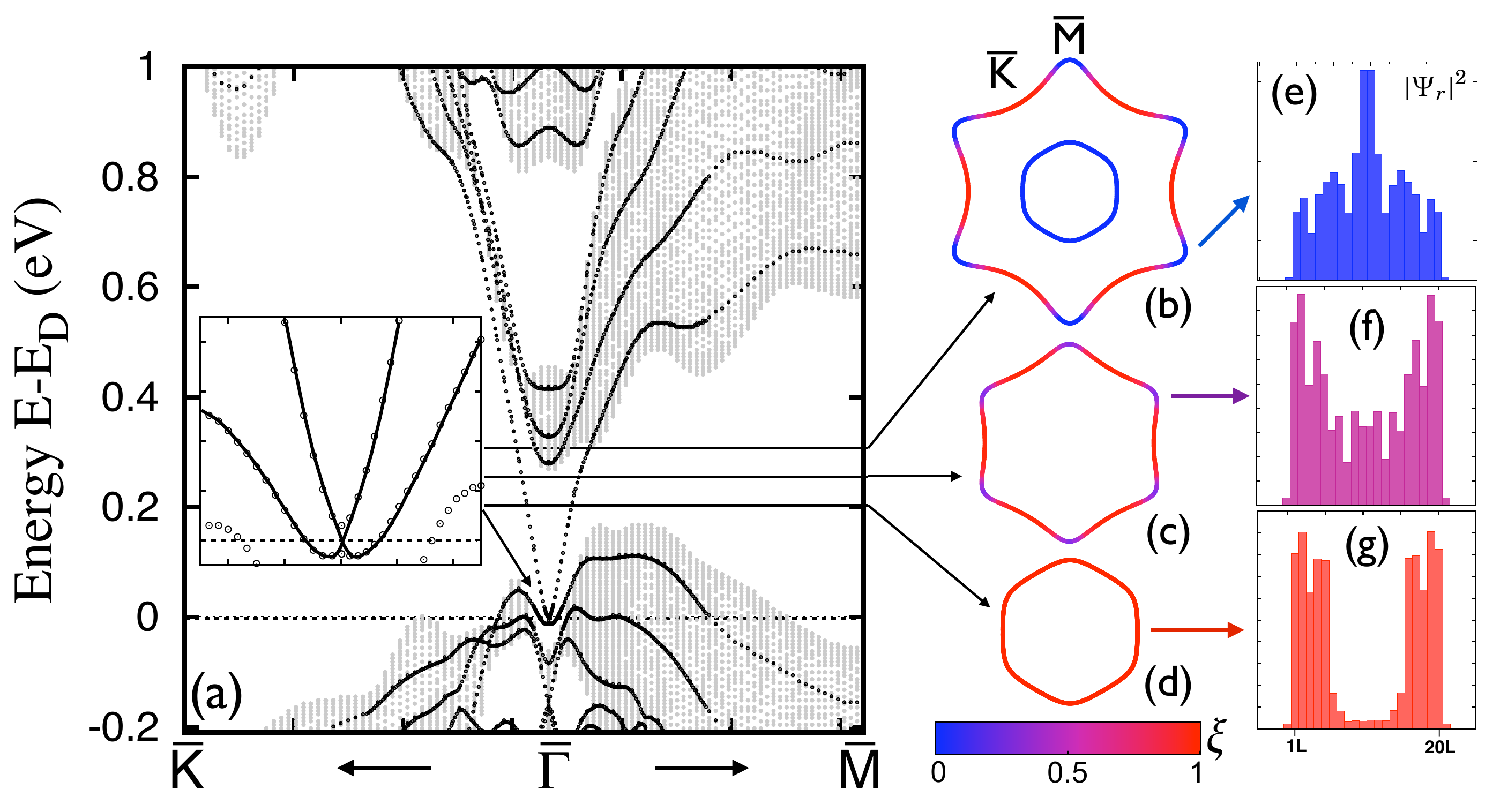}
\caption{\label{fig:1}(color online) (a) Band structure of the 30 layer $\text{Bi}_2\text{Te}_3$ slab (black dots) and surface projected 
bulk band structure (gray shaded areas). As an inset the surface Dirac state is highlighted. The Fermi surface at (b) \unit[316]{meV}, 
(c) \unit[270]{meV} and (d) \unit[205]{meV} above the Dirac energy $E_\text{D}$ is shown. Superimposed onto the Fermi surfaces is the 
probability density (e)-(g) of the surface states. Red color indicates a pure surface state mostly localized in the outermost quintuple layers, as shown 
in (g). Blue color refers to bulk-like contributions to the surface states along the $\bar{\Gamma}\bar{\text{M}}$ high symmetry line, as presented in (e) and (f).}
\end{figure*}
The calculated electronic bandstructure of the \BiTe surface (black dots) is shown together with the 
projected bulk band structure (gray shaded areas) along the hexagonal high symmetry lines in \f{1}(a). 
The findings are in good agreement with previous calculations \cite{Eremeev:2010,Yazyev:2010,Eremeev:2012}. 
The Dirac point of the \BiTe gapless surface state is located deep inside the bulk valence bands, at about 
$\unit[168]{meV}$ below the bulk valence band maximum (VBM), clearly reinforced by the distinct indirect bulk band gap. 
As is known from theory \cite{Park:2010} and experiment \cite{Liu:2010,Taskin:2012} the hybridization of the surface states 
localized on both sides of the slab leads to an artificial band gap opening of the Dirac state in the order of a few $\unit[]{meV}$. 
We observed a closure 
of the Dirac point band gap within an accuracy of $\unit[1]{meV}$ for 6 QL, albeit not showing any influence on the states character or the transport 
properties of the surface states.

With increasing Fermi level starting at the Dirac point the Fermi surface is circular, going to be hexagonal 
at $\approx\unit[150]{meV}$ (cf.~\fs{1}(c),(d)) and getting snowflake-like above $\unit[270]{meV}$ (\f{1}(b)). 
The origin of warping is the hybridization of the surface state with the bulk states in the 
$\bar{\Gamma}\bar{\text{M}}$ direction above $\unit[252]{meV}$ giving rise to a flat energy band, while in 
$\bar{\Gamma}\bar{\text{K}}$ direction the surface state stays isolated from the bulk states up to $\unit[0.8]{eV}$. The 
Fermi velocities in the high-symmetry directions differ remarkably, 
by  $\nicefrac{v_{\text{F}}^{\bar{\Gamma}\bar{\text{M}}}}{v_{\text{F}}^{\bar{\Gamma}\bar{\text{K}}}} \approx 0.4$, over a large energy range 
$E_{\text{F}}-E_{\text{D}}\approx \unit[0.25-0.5]{eV}$.

The DFT results serve as input to obtain the thermoelectric transport properties, using the layer-resolved 
transport distribution function (TDF) $\Sigma_{\|,i}(\mu)=\mathcal{L}_{\|,i}^{(0)}(\mu, 0)$ 
~\cite{Mahan:1996p508,Zahn:1998p15803}. The 
generalized conductance moments $\mathcal{L}_{\|,i}^{(n)}(\mu, T)$ are defined as 
\begin{eqnarray}
& \mathcal{L}_{\|,i}^{(n)}(\mu, T)= \\
&\frac{\tau_{\|}}{(2\pi)^2} \sum \limits_{\nu} \int d^2 \bm{k} \ |v^{\nu}_{\bm{k},(\|)}|^2 \ \mathcal{P}_{\bm{k}}^i \ (E^{\nu}_{\bm{k}} -\mu)^{n}\left( -\frac{\partial f_{(\mu,T)}}{\partial E} \right)_{E=E^{\nu}_{\bm{k}}} .\nonumber
\label{Tcoeff}
\end{eqnarray} 
$v^{\nu}_{\bm{k},(\|)}$ denote the group velocities in the directions of the hexagonal basal plane and $\mathcal{P}_{\bm{k}}^i$ is 
the layer-resolved probability amplitude of a Bloch state defined as
\begin{equation}
\int d\bm{r} |\mathring{\psi}_{\bm{k}}(\bm{r})|^2=\sum_i\mathcal{P}_{\bm{k}}^i=1\ .
\end{equation}
Here the relaxation time for \BiTe was fitted to experimental data and chosen to be constant 
in its absolute value $\tau=\unit[11]{fs}$ with respect to wave vector $\bm{k}$ and energy 
on the scale of $k_{B}T$ \cite{Hinsche:2011p15707}. No distinction of surface and bulk scattering was assumed 
to allow for a clear discussion of the electronic structure on the electronic 
transport. The influence of electron-phonon coupling was theoretically and experimentally found to be very weak and 
discussed more in detail in Appendix \ref{elph}.
Straightforward, the temperature- and doping-dependent in-plane 
electrical conductivity $\sigma_{_{\|}}$ and thermopower $S_{_{\|}}$ read as
\begin{eqnarray}
\sigma_{_{\|}}=2e^2 \mathcal{L}_{\|}^{(0)}(\mu, T) \quad \text{and} \quad S_{_{\|}}=\frac{1} {eT} \frac{\mathcal{L}_{\|}^{(1)}(\mu,T)} {\mathcal{L}_{\|}^{(0)}(\mu,T)}
\label{sigma}
\end{eqnarray}
for given chemical potential $\mu$ at temperature $T$ and carrier concentration $n$ 
\footnote{The fixed charge $N$ at varying $\mu(T)$ is determined by an integration 
 over the density of states $g(E)$ $N=\int \limits_{\mu-\Delta E}^{\text{VB}^{max}} \text{d}E \,  g(E) [f(\mu,T)-1]+
 \int \limits_{\text{CB}^{min}}^{\mu+\Delta E} \text{d}E \, g(E) f(\mu,T)$}.
All transport calculations presented below are performed with adaptive k-point mesh's 
larger than 500 points on a piece of the 2D Fermi surface which lies in the irreducible part of the 
Brillouin zone and at least 150 000 (56 million k-points) k-points in the entire 3D BZ for the bulk TDF at large (low) charge 
carrier concentrations. Detailed descriptions were given in earlier publications by Ref.~\onlinecite{Zahn:2011p15523,Yavorsky:2011p15466}. 
By means of the layer-resolved probability amplitude $\mathcal{P}_{\bm{k}}^i$ the thermoelectric transport properties 
can be decomposed into contributions of typical eigenstates. 
Within here, the distinction of surface states, characterized by a strong spatial localization in the outermost QL together 
with an exponential decay into the bulk and vacuum, and bulk states is required. This is done by probing 
a test eigenstate $\mathring{\psi}_{\bm{k}}(\bm{r})$ with the prototype surface eigenstate $\psi_{\bm{k},\text{SS}}(\bm{r})$ 
via $\xi_{\bm{k}}=\int d\bm{r} \psi^{*}_{\bm{k},\text{SS}}(\bm{r})\mathring{\psi}_{\bm{k}}(\bm{r})$. 
If $\xi_{\bm{k}}$ is larger than a given threshold close to one{\footnote{Throughout this study a threshold of 0.90 was used.}, 
the state $\bm{k}$ is considered a surface state. An example for the latter is given in \f{1}(g), by showing the layer-resolved 
probability amplitude. Most of the surface states probability amplitude is located at the outermost QL \cite{Henk:2012}. 
Furthermore, the information of the states character is shown on the Fermi surfaces \fs{1}(b)-(d). 
It is seen that all states of the surface band crossing the band gap posses surface state character with the corresponding 
spatial distribution (\f{1}(d)) up to an energy of about $\unit[220]{meV}$ above the Dirac point. 
From here, slight deviations of the state's prototype 
surface state character occur (cf. \f{1}(f)), as states in the $\bar{\Gamma}\bar{\text{M}}$ direction start to show 
hybridization between bulk and surface states. 
We note, that these changes start well below the bulk conduction band edge, $E_{\text{CBM}}-E_{\text{F}}\approx \unit[20]{meV}$. 
At elevated energies the states in $\bar{\Gamma}\bar{\text{M}}$ direction show clear bulk like character with a high probability amplitude 
in the center of the layer, e.g. shown in \f{1}(e). 
Consequently, $\xi_{\bm{k}} \rightarrow 0$ and these states 
do not behave as typical surface states, although originating 
from the Dirac band. As indicated earlier the surface states band is unaffected by hybridization effects in the 
$\bar{\Gamma}\bar{\text{K}}$ direction up to $E_{\text{F}}-E_{\text{D}}\approx \unit[0.8]{eV}$. 
As a result the Fermi surface show emerging bulk or surface character, most 
convincingly depicted in \f{1}(b). As will be discussed hereinafter, this fact leads to a strong influence on the 
electronic transport, exceptionally stating an almost constant electrical surface conductivity over a broad 
doping regime.

\begin{figure}[t]
\centering
\includegraphics[width=0.45\textwidth]{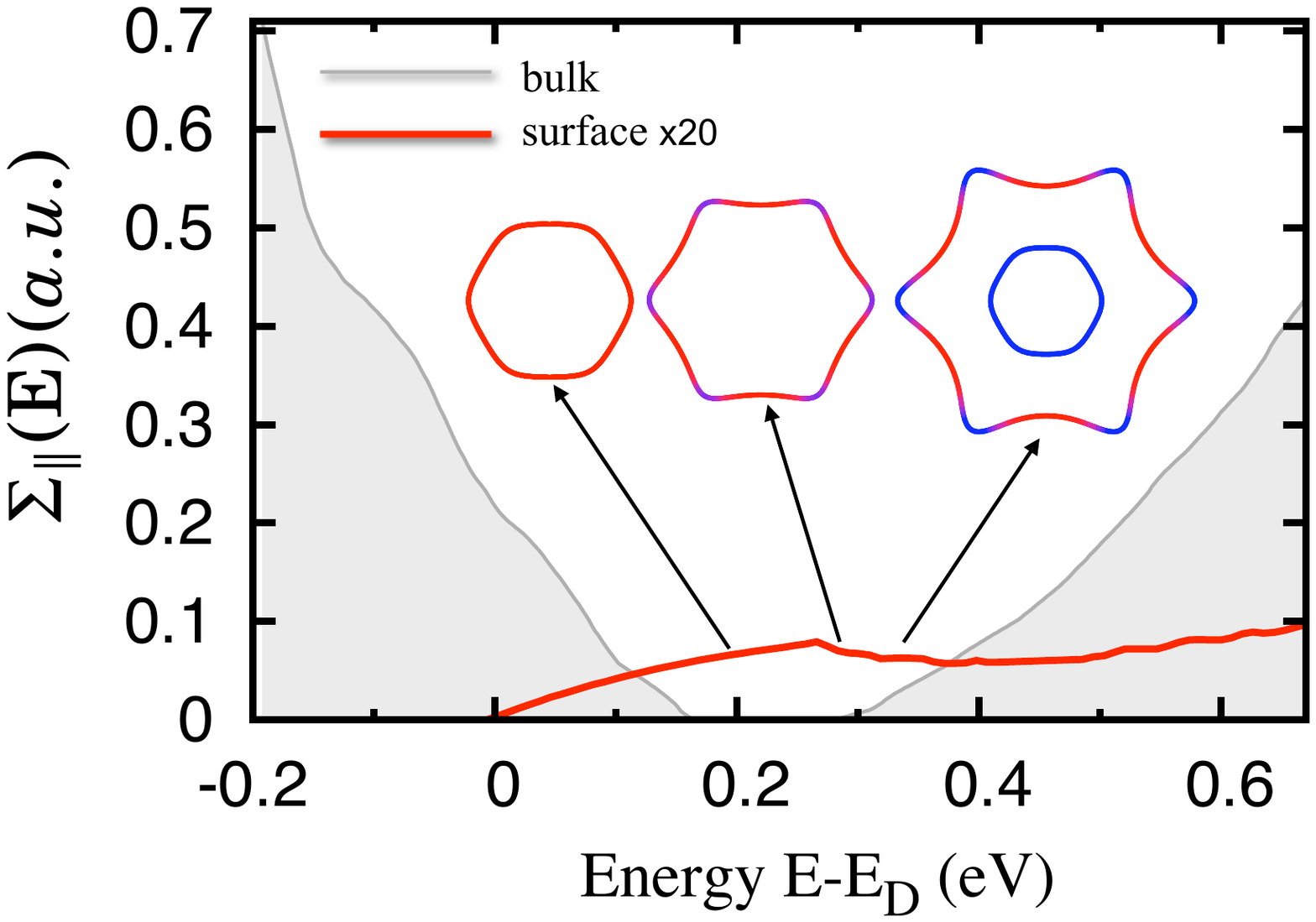}
\caption{\label{fig:2}(color online) Transport distribution function $\Sigma_{\|}(E)$ of the bulk and surface states of $\text{Bi}_2\text{Te}_3$. 
The red solid line refers to the contribution to $\Sigma_{\|}(E)$ of mainly the outermost quintuple layers (c.f. \f{1}(g)). The gray shaded area indicates 
the pure bulk contribution to $\Sigma_{\|}(E)$. Note that the surface contributions are multiplied by a factor of 20. 
The inset shows the corresponding Fermi surfaces as a function of energy (c.f. \f{1}(b)-(d)).}
\end{figure}
Basis of all transport properties discussed below is the TDF $\Sigma_{\|}(\mu)$, which can be understood as electrical 
conductivity at vanishing temperature. Due to hybridization, the TDF of the surface states 
contains contributions from surface and bulk states, which can be clearly separated using 
the projection technique by probability amplitudes as introduced above.
We will distinguish between contributions from the pure surface state (SS) (cf. \f{1}(g)), located at the surface, 
and bulk contributions to the surface states. The bulk TDF (gray shaded areas), as well as 
the surface (red solid line) contribution of the TDF are shown 
in \f{2}. The TDF of the surface contribution rises almost linearly with energy. 
Having in mind, that for a two-dimensional system the TDF scales as $\Sigma_{\|}(\mu) \propto \text{d}l_{\text{F}}\times v_{\|}$, 
whereas $\text{d}l_{\text{F}}$ is the length of the Fermi circle at chemical potential $\mu$, the linearity of the TDF 
in energy close to the Dirac point is obvious. Here, $v_{\|}$ is constant with energy, while $\text{d}l_{\text{F}}\propto E$. 
Small deviations from the latter arise about $E_{\text{F}}-E_{\text{D}}\approx \unit[0.1]{eV}$ and can be 
related to the hexagonal warping of the Fermi surface. 
At about $\unit[250]{meV}$ the increase of the TDF saturates and $\Sigma_{\|}^{\text{SS}}$ remains roughly 
constant over a wide range of energy.
%
As pointed out earlier, the pure surface states are not only spatially confined within approximately the outermost QL, but 
are also restricted to selected $\bm{k}$-directions because of the hybridization with the bulk 
states. In the ultimate limit of $\bm{k}$-selection only one Bloch function in $\bar{\Gamma}\bar{\text{K}}$ direction 
with a constant Fermi velocity $v_{\text{F}}^{\bar{\Gamma}\bar{\text{K}}}$ approximately realized for energies up to $\unit[0.8]{eV}$ above the 
Dirac point would be available for each surface state. As a consequence, the TDF of the surface state reads 
\begin{equation}
\Sigma_{\text{1D}}(\mu)=\frac{1}{\pi}\int \text{d}k \left(v_{\text{F}}^{\bar{\Gamma}\bar{\text{K}}}\right)^2 \delta(\mu-v_{\text{F}}^{\bar{\Gamma}\bar{\text{K}}}k)=\frac{L v_{\text{F}}^{\bar{\Gamma}\bar{\text{K}}}}{\pi}\ .
\end{equation}
Obviously, the transition from the two-dimensional character of the surface states into a one-dimensional 
one changes the energy dependence of the TDF from linear into constant (cf. \f{2}). With the TDF being directly related 
to $\mathcal{L}_{\|}^{(0)}(\mu, T)$ the electrical conductivity of the surface state is expected to 
be energy independent for electron doping as well.
%
We note, the fact of the Dirac point being buried deep inside the bulk valence bands causes 
a surface contribution to the TDF only for energies larger than $E_\text{D}$.
\begin{figure}[]
\centering
\includegraphics[width=0.49\textwidth]{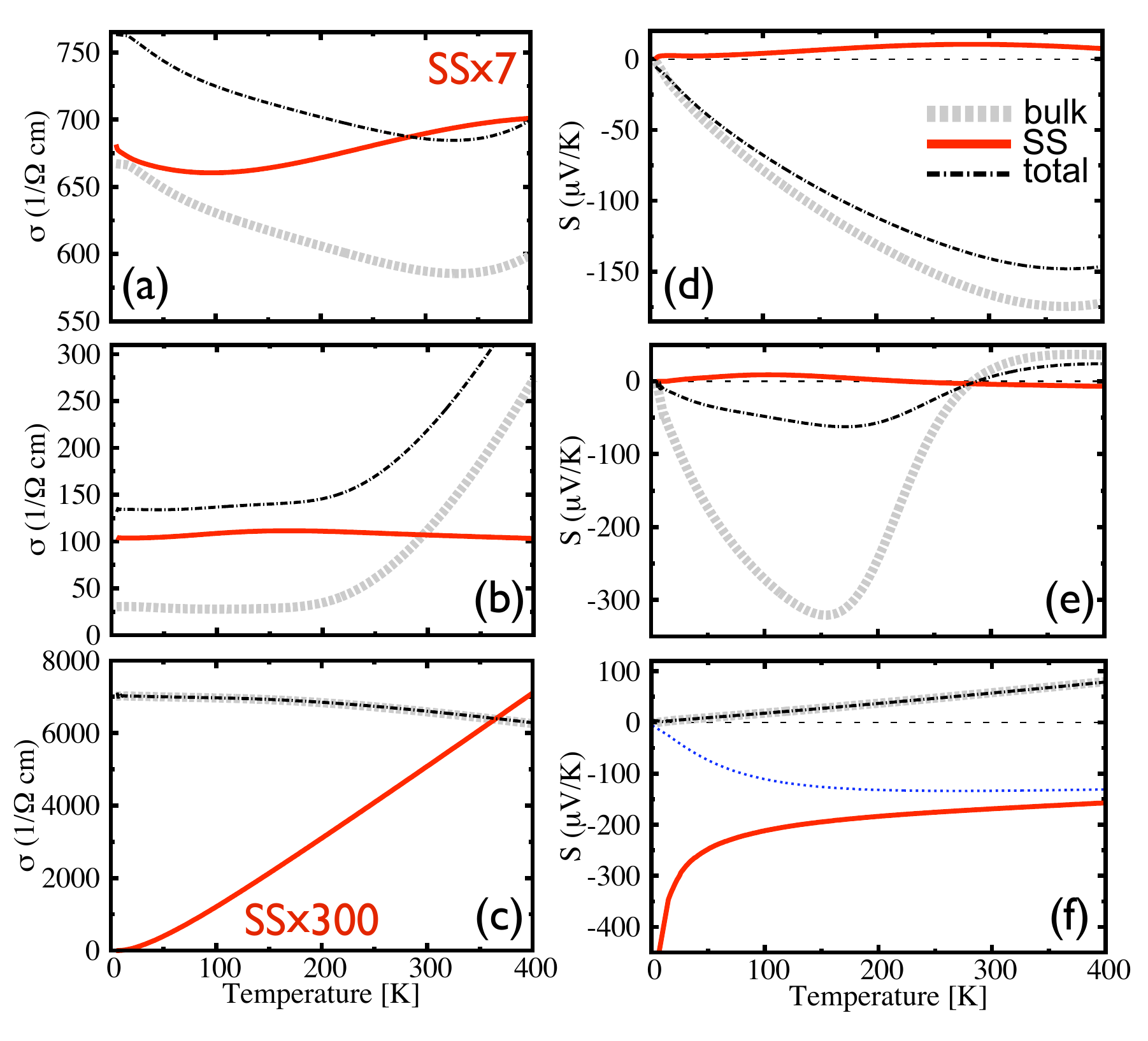}
\caption{\label{fig:3}(color online) Electrical conductivity (a)-(c) and thermopower (d)-(f) in dependence on temperature for three distinct 
charge carrier concentrations. (a) and (d) electron charge carrier concentration of $n = \unit[2 \times 10^{19}]{cm^{-3}}$, (b) and (e) 
electron charge carrier concentration of $n = \unit[1 \times 10^{18}]{cm^{-3}}$, (c) and (f) hole charge carrier concentration 
of $p = \unit[4 \times 10^{20}]{cm^{-3}}$. Pure bulk contributions are stated by gray dashed lines, the contribution of the surface 
states is given by red solid lines, while black dash-dotted lines show the total contribution of the half-infinite sample. In (f) the contribution of the surface state to the thermopower at $p = \unit[3 \times 10^{20}]{cm^{-3}}$ is given additionally (blue thin dotted line), to emphasize the thermodynamical limit of the thermopower at vanishing temperature. Further details can be found in the text.}
\end{figure}
In the following, we will discuss the doping- and temperature-dependent electrical conductivity and 
thermopower, as shown in \f{3}. As done before, we will distinguish between contributions 
from bulk states (gray dashed lines), surface states (red solid lines) and the total contribution (black dash-dotted line), 
defined as $\sigma_{\text{tot}}=\sigma_{\text{bulk}}+\sigma_{\text{SS}}$ and $S_{\text{tot}}=\nicefrac{(\sigma_{\text{bulk}}S_{\text{bulk}}+\sigma_{\text{SS}}S_{\text{SS}})}{\sigma_{\text{tot}}}$. 
Three typical charge carrier concentrations are chosen to reflect the overall behaviour of the transport properties. 
%
Due to the mere fact that a three-dimensional topological insulator offers robust metal-like surface states in the 
insulating bulk band gap, an enhanced electrical conductivity of 
the whole system is expected for very small charge carrier concentrations, i.e. the chemical potential 
being situated in the bulk band gap. The temperature dependence of the 
electrical conductivity for such a scenario, at an electron doping of $n=\unit[1 \times 10^{18}]{cm^{-3}}$, is shown in \f{3}(b). 
The surface contribution $\sigma_{\text{SS}}$ of the electrical conductivity is almost temperature independent 
at approximately $\unit[105]{(\Omega cm)^{-1}}$ for the entire temperature range. This behaviour is a 
consequence of the earlier discussed energy-independence of the TDF for electron doping. 
For low temperatures $\sigma_{\text{SS}}$ is up to 4 times larger than the bulk value of 
$\unit[28]{(\Omega cm)^{-1}}$ and the surface conductivity clearly dominates. 
For elevating temperatures bipolar bulk conduction leads to the well known 
exponential increase in $\sigma_{\text{bulk}}$ for narrow band gap semi-conductors 
and $\sigma_{\text{bulk}} > \sigma_{\text{SS}}$ holds for $T >\unit[300]{K}$. Due to the 
independence of $\sigma_{\text{SS}}$ on temperature the surface contribution to the total 
electrical conductivity is almost hidden and causes only a notable offset at low temperatures. 
However, such a behaviour was experimentally seen for thin \BiTe films \cite{Alpichshev:2010p15262,Taskin:2011}. 
A possibility to experimentally clarify whether surface states contribute to the total transport 
or not could be performed by measuring the total thermopower $S_{\text{tot}}$ of the system, 
which decomposed in it's parts is shown in \f{3}(e). 
For the bulk contribution of $S_{\text{tot}}$ the typical behaviour for a slightly doped 
narrow band gap bulk semiconductor is obtained. With increasing temperature the absolute value of $S_{\text{bulk}}$ 
rises linearly and the chemical potential shifts from the bulk conduction band into the bulk band 
gap. At a temperature of about $\unit[155]{K}$ 
large bulk values of the thermopower of $\unit[-320]{\mu V/K}$ are obtained. At higher temperatures 
$S_{\text{bulk}}$ saturates at the expected small values because of bipolar intrinsic transport. 
For a bulk semicondcutor in the intrinsic limit holds $S_{\text{max}}\sim \frac{E_{\text{G}}(T_{\text{max}})}{2T_{\text{max}}}$. 
The surface contribution $S_{\text{SS}}$ shows the expected metal-like behaviour with 
absolute values well below $\pm\unit[10]{\mu V/K}$. 
This leads in sum to a clearly diminished total thermopower. More precisely, assuming that $S_{\text{bulk}} \gg S_{\text{SS}}$, 
it is $S_{\text{tot}} \approx \nicefrac{S_{\text{bulk}}}{(\eta+1)}$ for $\eta = \nicefrac{\sigma_{\text{SS}}}{\sigma_{\text{bulk}}}$. 
The maximal absolute value of the total thermopower (black dashed-dotted line in \f{3}(e)) is found to be 
reduced to a fifth of the bulk value being $\unit[-62]{\mu V/K}$ at $\unit[170]{K}$, which corroborates 
the above-noted estimation. A clear transport contribution of the surface state will lead to a heavily 
decreased absolute value of the total thermopower. 

To complete the picture two additional doping regimes will be discussed. 
Usually bulk \BiTe is intrinsically electron-doped due to anti-site defects \cite{Hashibon:2011}. In \f{3}(a),(d) 
the electron doping amounts $n=\unit[2 \times 10^{19}]{cm^{-3}}$ and $\mu$ lies in 
or close to the bulk conduction band. Here $\sigma_{\text{tot}}$ is mainly defined 
by the bulk conductivity and reaches $\unit[685]{(\Omega cm)^{-1}}$ at room temperature. 
The surface contribution is about 12-15$\%$ for all temperatures, being almost constant with temperature, 
again reflecting the weak energy-dependence of the TDF. With the surface thermopower expectably behaving metal-like, 
but $\sigma_{\text{bulk}} \gg \sigma_{\text{SS}}$, the surface states impact on $S_{\text{tot}}$ 
is only moderate reducing $S_{\text{tot}}$  just up to 15$\%$ of the bulk value. 

Recent model calculations \cite{Ghaemi:2010,Takahashi:2012} for \BiTe proposed a dramatically enhanced 
thermoelectric powerfactor, i.e. $S\sigma^2$, at low temperatures 
and a location of the chemical potential near the Dirac point. 
To enter this regime, high hole doping rates are necessary. Here, we fixed the hole 
charge carrier concentration to $p=\unit[4 \times 10^{20}]{cm^{-3}}$ to discuss a possible 
enhancement of the thermoelectric powerfactor of \BiTe in the presence of gapless metal-like surface states. 
Indeed, as is shown in \f{3}(f), $S_{\text{SS}}$ possesses large semiconductor-like absolute values, 
even showing a divergence for $T\rightarrow 0$, with $\mu(T=0)$ coinciding with the Dirac energy $E_{\text{D}}$. 
This behaviour originates in the asymmetric slope 
of the TDF mocking a bulk band edge and yielding $S_{\text{SS}} \sim \nicefrac{-1}{(\mu-E_{\text{D}})}$
\footnote{We found additional small contributions to the surface state at energies 
$E \le E_{\text{D}}-\unit[250]{meV}$. The latter, do not noticeably contribute to the thermoelectric transport 
even at large hole charge carrier concentrations and high temperatures and can be omitted.}. 
Nevertheless the total thermopower and the bulk contribution to the thermopower are identical, both 
showing small positive values and a linear temperature dependence expected from a highly doped 
hole semicondcutor. The reason is the heavily suppressed contribution of the surface states 
due to the large difference between $\sigma_{\text{SS}}$ and $\sigma_{\text{bulk}}$ as shown in \f{3}(c). 
With the chemical potential deep in the bulk valence bands near the 
Dirac point, $\sigma_{\text{bulk}}$ dominates the total electrical transport 
by about a factor 300 and with this the contribution of $S_{\text{SS}}$ is negligible. 
If the Dirac point would be situated in the bulk band gap, i.e. available in \BiSe and \SbTex, 
the electronic thermoelectric transport is most probably enhanced with respect to bulk behaviour. In \BiTe 
such an enhancement is suppressed by the energetic position of the Dirac point, buried deep inside the bulk 
valence bands. 

We furthermore note, that the thermodynamical limit $S_{\text{SS}}(T \rightarrow 0) \rightarrow 0$ is reached as soon as 
the temperature dependent chemical potential at zero temperature is not identical to the Dirac energy, i.e. 
$\mu(T=0) \neq E_{\text{D}}$. This is emphasized in \f{3}(f) (blue thin dotted line) for a slightly smaller p-doping 
of  $p=\unit[3 \times 10^{20}]{cm^{-3}}$. Here, $\mu(T=0) = E_{\text{D}} + \unit[15]{meV}$ and the thermopower of the surface 
state vanishes at zero temperature. The total thermopower $S_{\text{tot}}$ of the thin film has always to vanish at zero temperature, 
regardless of the temperature dependence of the surface states' contribution (c.f. \fs{3}(d)-(f)).

In conclusion, we presented \textit{ab initio} calculations of the thermoelectric properties of \BiTe films. The contribution
of bulk and surface states to conductivity and thermopower are separated by a special projection technique. The contribution of 
the topological surface state is particularly pronounced in the low doping regime if the chemical potential lies in the 
bulk band gap. 

The conductivity of the semi-conductor \BiTe is enhanced by a constant contribution because of the surface state and reaches
values of a metallic system. The thermopower of bulk semi-conductors shows a pronounced maximum as a function of temperature.
The maximum value is determined by the size of the band gap. With the existence of the topological surface state this maximum 
value is drastically reduced towards metallic behaviour. A reduction of the total thermopower has been found in various 
experiments on thin film thermoelectric topological insulators, i.e. \BiTex, \SbTe and \BiSbTex, with, up to now, 
no clear explanation \cite{Peranio2006,Zastrow:2013p16357,Boulouz:2001p16385}. 

Consequently, the measured thermopower can be used to prove 
whether a surface state exists and contributes to the transport properties. 
To clearly distinguish between a topological surface state and a trivial surface state we suggest 
measurements of the thermopower at a single crystal. The 
contribution of the topological surface state is expected to be independent from the single 
crystal orientation with respect to the current direction since the topological SS occurs on all surfaces \cite{Lee:2009p16355}, 
while a trivial SS is restricted to selected surfaces. Following our discussion, a reduction to metallic 
behaviour of the thermopower and electrical conductivity is expected for all orientations of the single crystal.

\begin{acknowledgments}
  This work was supported by the Deutsche
  Forschungsgemeinschaft, SPP 1386 `Nanostrukturierte Thermoelektrika: 
  Theorie, Modellsysteme und kontrollierte Synthese'. N. F. Hinsche is
  member of the International Max Planck Research School for Science
  and Technology of Nanostructures. 
\end{acknowledgments} 

\appendix
\section{Electron-phonon scattering}
\label{elph}

Electron-phonon interactions are up to now not explicitly accounted for in our \textit{ab initio} calculations. The latter would most probably lead 
\textit{a priori} to a $1/T$ behaviour of \textit{only} the electron-phonon contribution to the bulk total relaxation time. Moreover we 
used the following ansatz for first estimations\footnote{Unfortunately, the precise numerical determination of the electron-phonon coupling on an \textit{ab initio} level is quite demanding for semi-conducting materials. The \textsc{\`{E}liashberg} function has to be determined for a rather large numbers of chemical potentials.}.

Applying \textsc{Matthiessens} rule the total electronic relaxation time, neglecting electron-electron and 
electron-magnon processes, reads as
\begin{equation}
\frac{1}{\tau}=\frac{1}{\tau_{el-imp}}+\frac{1}{\tau_{el-ph}}.
\end{equation}
Within our manuscript the total relaxation time $\tau$ was fitted to experimental transport measurements. In particular the 
thermopower in dependence of the electrical conductivity $S(\sigma)$ was analysed as first suggested by Ref.~\onlinecite{Scheidemantel:2003p14961}. 
The total relaxation time was found to be $\tau=\unit[11]{fs}$ for n and p-doped bulk materials, which is in good agreement to other theoretically $\tau=\unit[6-22]{fs}$ \cite{Scheidemantel:2003p14961,Park:2010p11006} and experimentally determined 
$\tau=\unit[36]{fs}$ relaxation times \cite{Stordeur:1988p15503}. More details can be found in a previous publication \cite{Hinsche:2011p15707}.

\begin{widetext}
Knowing the  \textsc{\`{E}liashberg} function $\alpha^{2}\mathcal{F}(\omega)$ allows to calculate the quasi-elastic electron-phonon scattering rate by \cite{Mahan:1990p16105}
\begin{eqnarray}
(\tau_{\text{e-ph}})^{-1}=
2\pi \int\limits_{0}^{\infty} \rm{d}\omega \; \alpha^{2}\mathcal{F}( \omega) 
\times \{ f^{0}(\epsilon_{F} +\hbar \omega) - f^{0}(\epsilon_{F} - \hbar \omega) + 2n^{0}(\omega) +1\} 
\overset{\epsilon_{F}=0}{=} 
4\pi \int\limits_{0}^{\infty} \rm{d}\omega \; \frac{\alpha^{2}\mathcal{F}(\omega)}{\sinh{\left( \frac{\hbar\omega}{k_{B}T}\right)}}
\label{for:eph_scatt}, 
\end{eqnarray}
\end{widetext}
with the coupling constant defined as
\begin{equation}
\lambda = 2 \int \frac{\alpha^{2}\mathcal{F}( \omega)}{\omega} d\omega
\label{for:lambda_mass}. 
\end{equation}
Here, $n^{0}$ and $f^{0}$ are the \textsc{Bose-Einstein} and \textsc{Fermi-Dirac} distribution, respectively. 
Once all phonon modes can contribute to the electron-phonon scattering, 
i.e $k_{B}T \geq \hbar\omega_{\text{max}}$ (for bulk \BiTex, i.e. $\omega_{\text{max}}\approx \unit[17]{meV}$), one readily obtains the relation 
$\tau_{\text{e-ph}}=\left( \frac{\hbar}{2\pi} \frac{1}{k_{B}\lambda}\right)\cdot \frac{1}{T}$ from eq.~\ref{for:eph_scatt}~\cite{Fabian:1999p16116}. 
Hence, in the high-temperature limit the scattering rate becomes linear in temperature, with a 
slope determined only by the integral value of the electron-phonon coupling constant $\lambda$. 

For strongly hole doped bulk Bi$_{2}$Te$_3$ ($p=8 \times \unit[10^{20}]{cm^{-3}}$, c.f. \f{4}, dotted lines) the coupling constant was estimated from the \textsc{Debye} temperature \cite{Shoemake:1969p16343} via the \textsc{McMillan} formula \cite{McMillan:1968p16342} to be $\lambda\approx 0.62$. 
However, for smaller, thermoelectric feasible, charge carrier concentrations (in the order of a few $\unit[10^{19}]{cm^{-3}}$) 
the coupling constant was very recently extracted from ARPES 
measurements \cite{Chen:2013p16344}. Very small bulk contributions 
of about $\lambda\approx 0.05$ ($\lambda\approx 0.17$) were found for p-doped (n-doped) samples \cite{Chen:2013p16344} (c.f. \f{4}, solid and dashed lines, respectively.). 
The electron-phonon coupling of the surface state is known to be very weak as well. Here, for \BiTe values of $\lambda_{SS}\approx 0.05$ 
could be revealed experimentally \cite{Chen:2013p16344} and theoretically \cite{Huang:2012p16206,Giraud:2012p16101}. 
These findings go along with previous similar results for the topological insulator \BiSe ($\lambda_{SS}\approx 0.076 \ldots 0.088$) \cite{Pan:2012p16341}. To shed some light on the possible influence of electron-phonon coupling to the total relaxation time, 
the latter is depicted in \f{4} for the previously described scenarios of the electron-phonon coupling parameter $\lambda$. 
\begin{figure}[]
\centering
\includegraphics[width=0.45\textwidth]{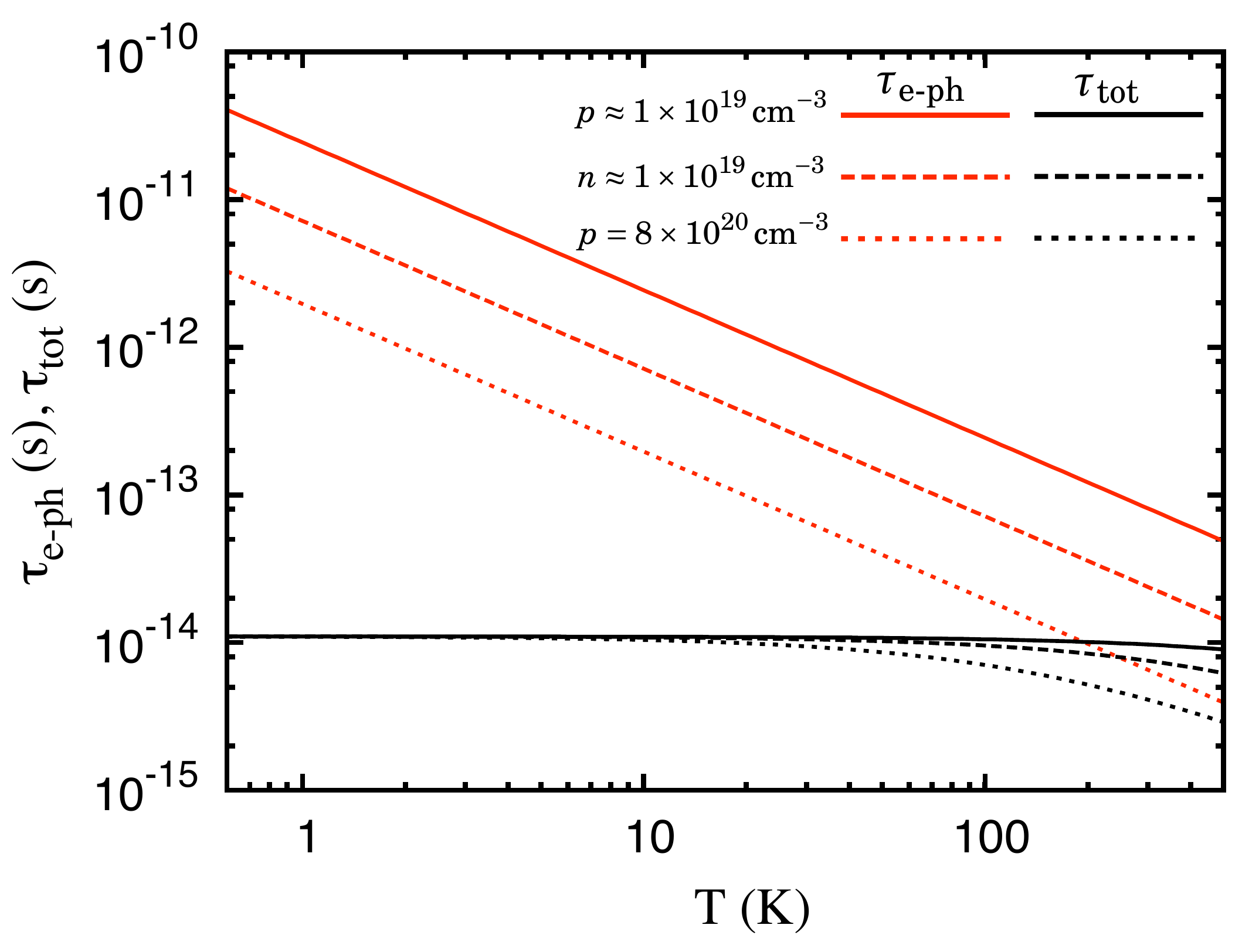}
\caption{\label{fig:4}(color online) Electron-phonon contribution (red lines) to the temperature-dependent total relaxation time (black lines) of \BiTe for different charge carrier concentrations and values of the integral electron-phonon coupling parameter $\lambda$. Further description within the text.}
\end{figure}

With the depicted dependence of $\tau_{\text{tot}}$ and $\tau_{\text{e-ph}}$, it is obvious that electron-phonon 
processes will not noticeably contribute to thermoelectric electron transport at room-temperature and below, as 
the electron-phonon scattering rates are at least one order of magnitude smaller, than contributions from electron-impurity 
scattering. Furthermore, with the bulk contribution of the electron-phonon coupling being clearly more weighted compared 
to the surface states' contribution, the previous discussed results on the topological surface states' signature on the 
thermoelectric transport still hold, while they might be even more pronounced as the ratio 
$\nicefrac{\sigma_{\text{SS}}}{\sigma_{\text{bulk}}}$ increases at higher temperatures.


 
\bibliography{paper.bbl}

\begin{thebibliography}{55}%
\makeatletter
\providecommand \@ifxundefined [1]{%
 \@ifx{#1\undefined}
}%
\providecommand \@ifnum [1]{%
 \ifnum #1\expandafter \@firstoftwo
 \else \expandafter \@secondoftwo
 \fi
}%
\providecommand \@ifx [1]{%
 \ifx #1\expandafter \@firstoftwo
 \else \expandafter \@secondoftwo
 \fi
}%
\providecommand \natexlab [1]{#1}%
\providecommand \enquote  [1]{``#1''}%
\providecommand \bibnamefont  [1]{#1}%
\providecommand \bibfnamefont [1]{#1}%
\providecommand \citenamefont [1]{#1}%
\providecommand \href@noop [0]{\@secondoftwo}%
\providecommand \href [0]{\begingroup \@sanitize@url \@href}%
\providecommand \@href[1]{\@@startlink{#1}\@@href}%
\providecommand \@@href[1]{\endgroup#1\@@endlink}%
\providecommand \@sanitize@url [0]{\catcode `\\12\catcode `\$12\catcode
  `\&12\catcode `\#12\catcode `\^12\catcode `\_12\catcode `\%12\relax}%
\providecommand \@@startlink[1]{}%
\providecommand \@@endlink[0]{}%
\providecommand \url  [0]{\begingroup\@sanitize@url \@url }%
\providecommand \@url [1]{\endgroup\@href {#1}{\urlprefix }}%
\providecommand \urlprefix  [0]{URL }%
\providecommand \Eprint [0]{\href }%
\@ifxundefined \urlstyle {%
  \providecommand \doi  [0]{\begingroup \@sanitize@url \@doi}%
  \providecommand \@doi [1]{\endgroup \@@startlink {\doibase
  #1}doi:\discretionary {}{}{}#1\@@endlink }%
}{%
  \providecommand \doi  [0]{doi:\discretionary{}{}{}\begingroup
  \urlstyle{rm}\Url }%
}%
\providecommand \doibase [0]{http://dx.doi.org/}%
\providecommand \Doi [0]{\begingroup \@sanitize@url \@Doi }%
\providecommand \@Doi  [1]{\endgroup\@@startlink{\doibase#1}\@@Doi}%
\providecommand \@@Doi [1]{#1\@@endlink}%
\providecommand \selectlanguage [0]{\@gobble}%
\providecommand \bibinfo  [0]{\@secondoftwo}%
\providecommand \bibfield  [0]{\@secondoftwo}%
\providecommand \translation [1]{[#1]}%
\providecommand \BibitemOpen [0]{}%
\providecommand \bibitemStop [0]{}%
\providecommand \bibitemNoStop [0]{.\EOS\space}%
\providecommand \EOS [0]{\spacefactor3000\relax}%
\providecommand \BibitemShut  [1]{\csname bibitem#1\endcsname}%
\bibitem [{\citenamefont {B{\"o}ttner}\ \emph {et~al.}(2006)\citenamefont
  {B{\"o}ttner}, \citenamefont {Chen},\ and\ \citenamefont
  {Venkatasubramanian}}]{Bottner:2006p2812}%
  \BibitemOpen
  \bibfield  {author} {\bibinfo {author} {\bibfnamefont {H.}~\bibnamefont
  {B{\"o}ttner}}, \bibinfo {author} {\bibfnamefont {G.}~\bibnamefont {Chen}}, \
  and\ \bibinfo {author} {\bibfnamefont {R.}~\bibnamefont
  {Venkatasubramanian}},\ }\Doi {10.1557/mrs2006.47} {\bibfield  {journal}
  {\bibinfo  {journal} {MRS bulletin},\ }\textbf {\bibinfo {volume} {31}},\
  \bibinfo {pages} {211} (\bibinfo {year} {2006})}\BibitemShut {NoStop}%
\bibitem [{\citenamefont {Snyder}\ and\ \citenamefont
  {Toberer}(2008)}]{Snyder:2008}%
  \BibitemOpen
  \bibfield  {author} {\bibinfo {author} {\bibfnamefont {G.~J.}\ \bibnamefont
  {Snyder}}\ and\ \bibinfo {author} {\bibfnamefont {E.~S.}\ \bibnamefont
  {Toberer}},\ }\Doi {10.1038/nmat2090} {\bibfield  {journal} {\bibinfo
  {journal} {Nature materials},\ }\textbf {\bibinfo {volume} {7}},\ \bibinfo
  {pages} {105} (\bibinfo {year} {2008})},\ ISSN \bibinfo {issn}
  {1476-1122}\BibitemShut {NoStop}%
\bibitem [{\citenamefont {Fu}\ \emph {et~al.}(2007)\citenamefont {Fu},
  \citenamefont {Kane},\ and\ \citenamefont {Mele}}]{Fu:2007p15433}%
  \BibitemOpen
  \bibfield  {author} {\bibinfo {author} {\bibfnamefont {L.}~\bibnamefont
  {Fu}}, \bibinfo {author} {\bibfnamefont {C.}~\bibnamefont {Kane}}, \ and\
  \bibinfo {author} {\bibfnamefont {E.}~\bibnamefont {Mele}},\ }\href@noop {}
  {\bibfield  {journal} {\bibinfo  {journal} {Phys. Rev. Lett.},\ }\textbf
  {\bibinfo {volume} {98}},\ \bibinfo {pages} {106803} (\bibinfo {year}
  {2007})}\BibitemShut {NoStop}%
\bibitem [{\citenamefont {Zhang}\ \emph {et~al.}(2009)\citenamefont {Zhang},
  \citenamefont {Liu}, \citenamefont {Qi}, \citenamefont {Dai}, \citenamefont
  {Fang},\ and\ \citenamefont {Zhang}}]{Zhang:2009}%
  \BibitemOpen
  \bibfield  {author} {\bibinfo {author} {\bibfnamefont {H.}~\bibnamefont
  {Zhang}}, \bibinfo {author} {\bibfnamefont {C.-X.}\ \bibnamefont {Liu}},
  \bibinfo {author} {\bibfnamefont {X.-L.}\ \bibnamefont {Qi}}, \bibinfo
  {author} {\bibfnamefont {X.}~\bibnamefont {Dai}}, \bibinfo {author}
  {\bibfnamefont {Z.}~\bibnamefont {Fang}}, \ and\ \bibinfo {author}
  {\bibfnamefont {S.-C.}\ \bibnamefont {Zhang}},\ }\Doi {10.1038/nphys1270}
  {\bibfield  {journal} {\bibinfo  {journal} {Nature Physics},\ }\textbf
  {\bibinfo {volume} {5}},\ \bibinfo {pages} {438} (\bibinfo {year} {2009})},\
  ISSN \bibinfo {issn} {1745-2473}\BibitemShut {NoStop}%
\bibitem [{\citenamefont {Hasan}\ and\ \citenamefont
  {Kane}(2010)}]{Hasan:2010p15275}%
  \BibitemOpen
  \bibfield  {author} {\bibinfo {author} {\bibfnamefont {M.}~\bibnamefont
  {Hasan}}\ and\ \bibinfo {author} {\bibfnamefont {C.}~\bibnamefont {Kane}},\
  }\Doi {10.1103/RevModPhys.82.3045} {\bibfield  {journal} {\bibinfo  {journal}
  {Reviews of Modern Physics},\ }\textbf {\bibinfo {volume} {82}},\ \bibinfo
  {pages} {3045} (\bibinfo {year} {2010})}\BibitemShut {NoStop}%
\bibitem [{\citenamefont {Mahan}(1989)}]{Mahan:1989}%
  \BibitemOpen
  \bibfield  {author} {\bibinfo {author} {\bibfnamefont {G.~D.}\ \bibnamefont
  {Mahan}},\ }\Doi {10.1063/1.342976} {\bibfield  {journal} {\bibinfo
  {journal} {J. Appl. Phys.},\ }\textbf {\bibinfo {volume} {4}},\ \bibinfo
  {pages} {1578} (\bibinfo {year} {1989})}\BibitemShut {NoStop}%
\bibitem [{\citenamefont {Hor}\ \emph {et~al.}(2009)\citenamefont {Hor},
  \citenamefont {Richardella}, \citenamefont {Roushan}, \citenamefont {Xia},
  \citenamefont {Checkelsky}, \citenamefont {Yazdani}, \citenamefont {Hasan},
  \citenamefont {Ong},\ and\ \citenamefont {Cava}}]{Hor:2009p15893}%
  \BibitemOpen
  \bibfield  {author} {\bibinfo {author} {\bibfnamefont {Y.}~\bibnamefont
  {Hor}}, \bibinfo {author} {\bibfnamefont {A.}~\bibnamefont {Richardella}},
  \bibinfo {author} {\bibfnamefont {P.}~\bibnamefont {Roushan}}, \bibinfo
  {author} {\bibfnamefont {Y.}~\bibnamefont {Xia}}, \bibinfo {author}
  {\bibfnamefont {J.}~\bibnamefont {Checkelsky}}, \bibinfo {author}
  {\bibfnamefont {A.}~\bibnamefont {Yazdani}}, \bibinfo {author} {\bibfnamefont
  {M.}~\bibnamefont {Hasan}}, \bibinfo {author} {\bibfnamefont
  {N.}~\bibnamefont {Ong}}, \ and\ \bibinfo {author} {\bibfnamefont
  {R.}~\bibnamefont {Cava}},\ }\Doi {10.1103/PhysRevB.79.195208} {\bibfield
  {journal} {\bibinfo  {journal} {Physical Review B},\ }\textbf {\bibinfo
  {volume} {79}},\ \bibinfo {pages} {195208} (\bibinfo {year}
  {2009})}\BibitemShut {NoStop}%
\bibitem [{\citenamefont {Ghaemi}\ \emph {et~al.}(2010)\citenamefont {Ghaemi},
  \citenamefont {Mong},\ and\ \citenamefont {Moore}}]{Ghaemi:2010}%
  \BibitemOpen
  \bibfield  {author} {\bibinfo {author} {\bibfnamefont {P.}~\bibnamefont
  {Ghaemi}}, \bibinfo {author} {\bibfnamefont {R.}~\bibnamefont {Mong}}, \ and\
  \bibinfo {author} {\bibfnamefont {J.}~\bibnamefont {Moore}},\ }\Doi
  {10.1103/PhysRevLett.105.166603} {\bibfield  {journal} {\bibinfo  {journal}
  {Physical Review Letters},\ }\textbf {\bibinfo {volume} {105}},\ \bibinfo
  {pages} {1} (\bibinfo {year} {2010})}\BibitemShut {NoStop}%
\bibitem [{\citenamefont {Takahashi}\ and\ \citenamefont
  {Murakami}(2012)}]{Takahashi:2012}%
  \BibitemOpen
  \bibfield  {author} {\bibinfo {author} {\bibfnamefont {R.}~\bibnamefont
  {Takahashi}}\ and\ \bibinfo {author} {\bibfnamefont {S.}~\bibnamefont
  {Murakami}},\ }\Doi {10.1088/0268-1242/27/12/124005} {\bibfield  {journal}
  {\bibinfo  {journal} {Semiconductor Science and Technology},\ }\textbf
  {\bibinfo {volume} {27}},\ \bibinfo {pages} {124005} (\bibinfo {year}
  {2012})},\ ISSN \bibinfo {issn} {0268-1242}\BibitemShut {NoStop}%
\bibitem [{\citenamefont {Henk}\ \emph
  {et~al.}(2012){\natexlab{a}}\citenamefont {Henk}, \citenamefont {Flieger},
  \citenamefont {Maznichenko}, \citenamefont {Mertig}, \citenamefont {Ernst},
  \citenamefont {Eremeev},\ and\ \citenamefont {Chulkov}}]{Henk:2012p16289}%
  \BibitemOpen
  \bibfield  {author} {\bibinfo {author} {\bibfnamefont {J.}~\bibnamefont
  {Henk}}, \bibinfo {author} {\bibfnamefont {M.}~\bibnamefont {Flieger}},
  \bibinfo {author} {\bibfnamefont {I.~V.}\ \bibnamefont {Maznichenko}},
  \bibinfo {author} {\bibfnamefont {I.}~\bibnamefont {Mertig}}, \bibinfo
  {author} {\bibfnamefont {A.}~\bibnamefont {Ernst}}, \bibinfo {author}
  {\bibfnamefont {S.~V.}\ \bibnamefont {Eremeev}}, \ and\ \bibinfo {author}
  {\bibfnamefont {E.~V.}\ \bibnamefont {Chulkov}},\ }\Doi
  {10.1103/PhysRevLett.109.076801} {\bibfield  {journal} {\bibinfo  {journal}
  {Phys. Rev. Lett.},\ }\textbf {\bibinfo {volume} {109}},\ \bibinfo {pages}
  {076801} (\bibinfo {year} {2012}{\natexlab{a}})}\BibitemShut {NoStop}%
\bibitem [{\citenamefont {Ando}(2013)}]{Ando2013}%
  \BibitemOpen
  \bibfield  {author} {\bibinfo {author} {\bibfnamefont {Y.}~\bibnamefont
  {Ando}},\ }\href@noop {} {\bibfield  {journal} {\bibinfo  {journal} {J. Phys.
  Soc. Jpn.},\ }\textbf {\bibinfo {volume} {82}},\ \bibinfo {pages} {102001}
  (\bibinfo {year} {2013})}\BibitemShut {NoStop}%
\bibitem [{\citenamefont {Qu}\ \emph {et~al.}(2010)\citenamefont {Qu},
  \citenamefont {Hor}, \citenamefont {Xiong}, \citenamefont {Cava},\ and\
  \citenamefont {Ong}}]{Qu:2010}%
  \BibitemOpen
  \bibfield  {author} {\bibinfo {author} {\bibfnamefont {D.-X.}\ \bibnamefont
  {Qu}}, \bibinfo {author} {\bibfnamefont {Y.~S.}\ \bibnamefont {Hor}},
  \bibinfo {author} {\bibfnamefont {J.}~\bibnamefont {Xiong}}, \bibinfo
  {author} {\bibfnamefont {R.~J.}\ \bibnamefont {Cava}}, \ and\ \bibinfo
  {author} {\bibfnamefont {N.~P.}\ \bibnamefont {Ong}},\ }\Doi
  {10.1126/science.1189792} {\bibfield  {journal} {\bibinfo  {journal} {Science
  (New York, N.Y.)},\ }\textbf {\bibinfo {volume} {329}},\ \bibinfo {pages}
  {821} (\bibinfo {year} {2010})},\ ISSN \bibinfo {issn}
  {1095-9203}\BibitemShut {NoStop}%
\bibitem [{\citenamefont {Taskin}\ \emph
  {et~al.}(2011){\natexlab{a}}\citenamefont {Taskin}, \citenamefont {Ren},
  \citenamefont {Sasaki}, \citenamefont {Segawa},\ and\ \citenamefont
  {Ando}}]{Taskin:2011p15963}%
  \BibitemOpen
  \bibfield  {author} {\bibinfo {author} {\bibfnamefont {A.}~\bibnamefont
  {Taskin}}, \bibinfo {author} {\bibfnamefont {Z.}~\bibnamefont {Ren}},
  \bibinfo {author} {\bibfnamefont {S.}~\bibnamefont {Sasaki}}, \bibinfo
  {author} {\bibfnamefont {K.}~\bibnamefont {Segawa}}, \ and\ \bibinfo {author}
  {\bibfnamefont {Y.}~\bibnamefont {Ando}},\ }\href@noop {} {\bibfield
  {journal} {\bibinfo  {journal} {Phys. Rev. Lett.},\ }\textbf {\bibinfo
  {volume} {107}},\ \bibinfo {pages} {16801} (\bibinfo {year}
  {2011}{\natexlab{a}})}\BibitemShut {NoStop}%
\bibitem [{\citenamefont {Mertig}(1999)}]{Mertig:1999p12776}%
  \BibitemOpen
  \bibfield  {author} {\bibinfo {author} {\bibfnamefont {I.}~\bibnamefont
  {Mertig}},\ }\Doi {10.1088/0034-4885/62/2/004} {\bibfield  {journal}
  {\bibinfo  {journal} {Reports on Progress in Physics},\ }\textbf {\bibinfo
  {volume} {62}},\ \bibinfo {pages} {237} (\bibinfo {year} {1999})}\BibitemShut
  {NoStop}%
\bibitem [{\citenamefont {Hinsche}\ \emph
  {et~al.}(2011){\natexlab{a}}\citenamefont {Hinsche}, \citenamefont {Mertig},\
  and\ \citenamefont {Zahn}}]{Hinsche:2011p15276}%
  \BibitemOpen
  \bibfield  {author} {\bibinfo {author} {\bibfnamefont {N.}~\bibnamefont
  {Hinsche}}, \bibinfo {author} {\bibfnamefont {I.}~\bibnamefont {Mertig}}, \
  and\ \bibinfo {author} {\bibfnamefont {P.}~\bibnamefont {Zahn}},\ }\Doi
  {10.1088/0953-8984/23/29/295502} {\bibfield  {journal} {\bibinfo  {journal}
  {J. Phys.: Condens. Matter},\ }\textbf {\bibinfo {volume} {23}},\ \bibinfo
  {pages} {295502} (\bibinfo {year} {2011}{\natexlab{a}})}\BibitemShut
  {NoStop}%
\bibitem [{\citenamefont {Thonhauser}\ \emph {et~al.}(2003)\citenamefont
  {Thonhauser}, \citenamefont {Scheidemantel}, \citenamefont {Sofo},
  \citenamefont {Badding},\ and\ \citenamefont
  {Mahan}}]{Thonhauser:2003p14996}%
  \BibitemOpen
  \bibfield  {author} {\bibinfo {author} {\bibfnamefont {T.}~\bibnamefont
  {Thonhauser}}, \bibinfo {author} {\bibfnamefont {T.}~\bibnamefont
  {Scheidemantel}}, \bibinfo {author} {\bibfnamefont {J.}~\bibnamefont {Sofo}},
  \bibinfo {author} {\bibfnamefont {J.}~\bibnamefont {Badding}}, \ and\
  \bibinfo {author} {\bibfnamefont {G.}~\bibnamefont {Mahan}},\ }\Doi
  {10.1103/PhysRevB.68.085201} {\bibfield  {journal} {\bibinfo  {journal}
  {Physical Review B},\ }\textbf {\bibinfo {volume} {68}},\ \bibinfo {pages}
  {085201} (\bibinfo {year} {2003})}\BibitemShut {NoStop}%
\bibitem [{\citenamefont {Huang}\ and\ \citenamefont
  {Kaviany}(2008)}]{Huang:2008p559}%
  \BibitemOpen
  \bibfield  {author} {\bibinfo {author} {\bibfnamefont {B.}~\bibnamefont
  {Huang}}\ and\ \bibinfo {author} {\bibfnamefont {M.}~\bibnamefont
  {Kaviany}},\ }\href@noop {} {\bibfield  {journal} {\bibinfo  {journal}
  {Physical Review B},\ }\textbf {\bibinfo {volume} {77}},\ \bibinfo {pages}
  {125209} (\bibinfo {year} {2008})}\BibitemShut {NoStop}%
\bibitem [{\citenamefont {Park}\ \emph
  {et~al.}(2010){\natexlab{a}}\citenamefont {Park}, \citenamefont {Song},
  \citenamefont {Medvedeva}, \citenamefont {Kim}, \citenamefont {Kim},\ and\
  \citenamefont {Freeman}}]{Park:2010p11006}%
  \BibitemOpen
  \bibfield  {author} {\bibinfo {author} {\bibfnamefont {M.~S.}\ \bibnamefont
  {Park}}, \bibinfo {author} {\bibfnamefont {J.-H.}\ \bibnamefont {Song}},
  \bibinfo {author} {\bibfnamefont {J.~E.}\ \bibnamefont {Medvedeva}}, \bibinfo
  {author} {\bibfnamefont {M.}~\bibnamefont {Kim}}, \bibinfo {author}
  {\bibfnamefont {I.~G.}\ \bibnamefont {Kim}}, \ and\ \bibinfo {author}
  {\bibfnamefont {A.~J.}\ \bibnamefont {Freeman}},\ }\href@noop {} {\bibfield
  {journal} {\bibinfo  {journal} {Phys. Rev. B},\ }\textbf {\bibinfo {volume}
  {81}},\ \bibinfo {pages} {155211} (\bibinfo {year}
  {2010}{\natexlab{a}})}\BibitemShut {NoStop}%
\bibitem [{\citenamefont {Hinsche}\ \emph {et~al.}(2012)\citenamefont
  {Hinsche}, \citenamefont {Yavorsky}, \citenamefont {Gradhand}, \citenamefont
  {Czerner}, \citenamefont {Winkler}, \citenamefont {K{\"o}nig}, \citenamefont
  {B{\"o}ttner}, \citenamefont {Mertig},\ and\ \citenamefont
  {Zahn}}]{Hinsche:2012p16003}%
  \BibitemOpen
  \bibfield  {author} {\bibinfo {author} {\bibfnamefont {N.}~\bibnamefont
  {Hinsche}}, \bibinfo {author} {\bibfnamefont {B.}~\bibnamefont {Yavorsky}},
  \bibinfo {author} {\bibfnamefont {M.}~\bibnamefont {Gradhand}}, \bibinfo
  {author} {\bibfnamefont {M.}~\bibnamefont {Czerner}}, \bibinfo {author}
  {\bibfnamefont {M.}~\bibnamefont {Winkler}}, \bibinfo {author} {\bibfnamefont
  {J.}~\bibnamefont {K{\"o}nig}}, \bibinfo {author} {\bibfnamefont
  {H.}~\bibnamefont {B{\"o}ttner}}, \bibinfo {author} {\bibfnamefont
  {I.}~\bibnamefont {Mertig}}, \ and\ \bibinfo {author} {\bibfnamefont
  {P.}~\bibnamefont {Zahn}},\ }\Doi {10.1103/PhysRevB.86.085323} {\bibfield
  {journal} {\bibinfo  {journal} {Physical Review B},\ }\textbf {\bibinfo
  {volume} {86}},\ \bibinfo {pages} {085323} (\bibinfo {year}
  {2012})}\BibitemShut {NoStop}%
\bibitem [{\citenamefont {Hinsche}\ \emph
  {et~al.}(2011){\natexlab{b}}\citenamefont {Hinsche}, \citenamefont
  {Yavorsky}, \citenamefont {Mertig},\ and\ \citenamefont
  {Zahn}}]{Hinsche:2011p15707}%
  \BibitemOpen
  \bibfield  {author} {\bibinfo {author} {\bibfnamefont {N.}~\bibnamefont
  {Hinsche}}, \bibinfo {author} {\bibfnamefont {B.}~\bibnamefont {Yavorsky}},
  \bibinfo {author} {\bibfnamefont {I.}~\bibnamefont {Mertig}}, \ and\ \bibinfo
  {author} {\bibfnamefont {P.}~\bibnamefont {Zahn}},\ }\href@noop {} {\bibfield
   {journal} {\bibinfo  {journal} {Physical Review B},\ }\textbf {\bibinfo
  {volume} {84}},\ \bibinfo {pages} {165214} (\bibinfo {year}
  {2011}{\natexlab{b}})}\BibitemShut {NoStop}%
\bibitem [{\citenamefont {Gradhand}\ \emph {et~al.}(2009)\citenamefont
  {Gradhand}, \citenamefont {Czerner}, \citenamefont {Fedorov}, \citenamefont
  {Zahn}, \citenamefont {Yavorsky}, \citenamefont {Szunyogh},\ and\
  \citenamefont {Mertig}}]{Gradhand:2009p7460}%
  \BibitemOpen
  \bibfield  {author} {\bibinfo {author} {\bibfnamefont {M.}~\bibnamefont
  {Gradhand}}, \bibinfo {author} {\bibfnamefont {M.}~\bibnamefont {Czerner}},
  \bibinfo {author} {\bibfnamefont {D.~V.}\ \bibnamefont {Fedorov}}, \bibinfo
  {author} {\bibfnamefont {P.}~\bibnamefont {Zahn}}, \bibinfo {author}
  {\bibfnamefont {B.~Y.}\ \bibnamefont {Yavorsky}}, \bibinfo {author}
  {\bibfnamefont {L.}~\bibnamefont {Szunyogh}}, \ and\ \bibinfo {author}
  {\bibfnamefont {I.}~\bibnamefont {Mertig}},\ }\Doi
  {10.1103/PhysRevB.80.224413} {\bibfield  {journal} {\bibinfo  {journal}
  {Phys. Rev. B},\ }\textbf {\bibinfo {volume} {80}},\ \bibinfo {pages}
  {224413} (\bibinfo {year} {2009})}\BibitemShut {NoStop}%
\bibitem [{\citenamefont {Vosko}\ and\ \citenamefont {Wilk}(1980)}]{Vosko1980}%
  \BibitemOpen
  \bibfield  {author} {\bibinfo {author} {\bibfnamefont {S.~H.}\ \bibnamefont
  {Vosko}}\ and\ \bibinfo {author} {\bibfnamefont {L.}~\bibnamefont {Wilk}},\
  }\Doi {10.1103/PhysRevB.22.3812} {\bibfield  {journal} {\bibinfo  {journal}
  {Phys. Rev. B},\ }\textbf {\bibinfo {volume} {22}},\ \bibinfo {pages} {3812}
  (\bibinfo {year} {1980})}\BibitemShut {NoStop}%
\bibitem [{Lan(1998)}]{Landolt}%
  \BibitemOpen
  \enquote {\bibinfo {title} {{Landolt-B{\"o}rnstein New Series, group
  III/41C}},}\ \ (\bibinfo  {publisher} {Springer Verlag},\ \bibinfo {address}
  {Berlin},\ \bibinfo {year} {1998})\BibitemShut {NoStop}%
\bibitem [{\citenamefont {Eremeev}\ \emph {et~al.}(2010)\citenamefont
  {Eremeev}, \citenamefont {Koroteev},\ and\ \citenamefont
  {Chulkov}}]{Eremeev:2010}%
  \BibitemOpen
  \bibfield  {author} {\bibinfo {author} {\bibfnamefont {S.~V.}\ \bibnamefont
  {Eremeev}}, \bibinfo {author} {\bibfnamefont {Y.~M.}\ \bibnamefont
  {Koroteev}}, \ and\ \bibinfo {author} {\bibfnamefont {E.~V.}\ \bibnamefont
  {Chulkov}},\ }\Doi {10.1134/S0021364010080059} {\bibfield  {journal}
  {\bibinfo  {journal} {JETP Letters},\ }\textbf {\bibinfo {volume} {91}},\
  \bibinfo {pages} {387} (\bibinfo {year} {2010})},\ ISSN \bibinfo {issn}
  {0021-3640}\BibitemShut {NoStop}%
\bibitem [{\citenamefont {Yazyev}\ \emph {et~al.}(2010)\citenamefont {Yazyev},
  \citenamefont {Moore},\ and\ \citenamefont {Louie}}]{Yazyev:2010}%
  \BibitemOpen
  \bibfield  {author} {\bibinfo {author} {\bibfnamefont {O.}~\bibnamefont
  {Yazyev}}, \bibinfo {author} {\bibfnamefont {J.}~\bibnamefont {Moore}}, \
  and\ \bibinfo {author} {\bibfnamefont {S.}~\bibnamefont {Louie}},\ }\Doi
  {10.1103/PhysRevLett.105.266806} {\bibfield  {journal} {\bibinfo  {journal}
  {Physical Review Letters},\ }\textbf {\bibinfo {volume} {105}},\ \bibinfo
  {pages} {1} (\bibinfo {year} {2010})},\ ISSN \bibinfo {issn}
  {0031-9007}\BibitemShut {NoStop}%
\bibitem [{\citenamefont {Eremeev}\ \emph {et~al.}(2012)\citenamefont
  {Eremeev}, \citenamefont {Landolt}, \citenamefont {Menshchikova},
  \citenamefont {Slomski}, \citenamefont {Koroteev}, \citenamefont {Aliev},
  \citenamefont {Babanly}, \citenamefont {Henk}, \citenamefont {Ernst},
  \citenamefont {Patthey}, \citenamefont {Eich}, \citenamefont {Khajetoorians},
  \citenamefont {Hagemeister}, \citenamefont {Pietzsch}, \citenamefont {Wiebe},
  \citenamefont {Wiesendanger}, \citenamefont {Echenique}, \citenamefont
  {Tsirkin}, \citenamefont {Amiraslanov}, \citenamefont {Dil},\ and\
  \citenamefont {Chulkov}}]{Eremeev:2012}%
  \BibitemOpen
  \bibfield  {author} {\bibinfo {author} {\bibfnamefont {S.~V.}\ \bibnamefont
  {Eremeev}}, \bibinfo {author} {\bibfnamefont {G.}~\bibnamefont {Landolt}},
  \bibinfo {author} {\bibfnamefont {T.~V.}\ \bibnamefont {Menshchikova}},
  \bibinfo {author} {\bibfnamefont {B.}~\bibnamefont {Slomski}}, \bibinfo
  {author} {\bibfnamefont {Y.~M.}\ \bibnamefont {Koroteev}}, \bibinfo {author}
  {\bibfnamefont {Z.~S.}\ \bibnamefont {Aliev}}, \bibinfo {author}
  {\bibfnamefont {M.~B.}\ \bibnamefont {Babanly}}, \bibinfo {author}
  {\bibfnamefont {J.}~\bibnamefont {Henk}}, \bibinfo {author} {\bibfnamefont
  {A.}~\bibnamefont {Ernst}}, \bibinfo {author} {\bibfnamefont
  {L.}~\bibnamefont {Patthey}}, \bibinfo {author} {\bibfnamefont
  {A.}~\bibnamefont {Eich}}, \bibinfo {author} {\bibfnamefont {A.~A.}\
  \bibnamefont {Khajetoorians}}, \bibinfo {author} {\bibfnamefont
  {J.}~\bibnamefont {Hagemeister}}, \bibinfo {author} {\bibfnamefont
  {O.}~\bibnamefont {Pietzsch}}, \bibinfo {author} {\bibfnamefont
  {J.}~\bibnamefont {Wiebe}}, \bibinfo {author} {\bibfnamefont
  {R.}~\bibnamefont {Wiesendanger}}, \bibinfo {author} {\bibfnamefont {P.~M.}\
  \bibnamefont {Echenique}}, \bibinfo {author} {\bibfnamefont {S.~S.}\
  \bibnamefont {Tsirkin}}, \bibinfo {author} {\bibfnamefont {I.~R.}\
  \bibnamefont {Amiraslanov}}, \bibinfo {author} {\bibfnamefont {J.~H.}\
  \bibnamefont {Dil}}, \ and\ \bibinfo {author} {\bibfnamefont {E.~V.}\
  \bibnamefont {Chulkov}},\ }\Doi {10.1038/ncomms1638} {\bibfield  {journal}
  {\bibinfo  {journal} {Nature communications},\ }\textbf {\bibinfo {volume}
  {3}},\ \bibinfo {pages} {635} (\bibinfo {year} {2012})},\ ISSN \bibinfo
  {issn} {2041-1723}\BibitemShut {NoStop}%
\bibitem [{\citenamefont {Park}\ \emph
  {et~al.}(2010){\natexlab{b}}\citenamefont {Park}, \citenamefont {Heremans},
  \citenamefont {Scarola},\ and\ \citenamefont {Minic}}]{Park:2010}%
  \BibitemOpen
  \bibfield  {author} {\bibinfo {author} {\bibfnamefont {K.}~\bibnamefont
  {Park}}, \bibinfo {author} {\bibfnamefont {J.}~\bibnamefont {Heremans}},
  \bibinfo {author} {\bibfnamefont {V.}~\bibnamefont {Scarola}}, \ and\
  \bibinfo {author} {\bibfnamefont {D.}~\bibnamefont {Minic}},\ }\Doi
  {10.1103/PhysRevLett.105.186801} {\bibfield  {journal} {\bibinfo  {journal}
  {Physical Review Letters},\ }\textbf {\bibinfo {volume} {105}},\ \bibinfo
  {pages} {1} (\bibinfo {year} {2010}{\natexlab{b}})},\ ISSN \bibinfo {issn}
  {0031-9007}\BibitemShut {NoStop}%
\bibitem [{\citenamefont {Liu}\ \emph {et~al.}(2010)\citenamefont {Liu},
  \citenamefont {Zhang}, \citenamefont {Yan}, \citenamefont {Qi}, \citenamefont
  {Frauenheim}, \citenamefont {Dai}, \citenamefont {Fang},\ and\ \citenamefont
  {Zhang}}]{Liu:2010}%
  \BibitemOpen
  \bibfield  {author} {\bibinfo {author} {\bibfnamefont {C.-X.}\ \bibnamefont
  {Liu}}, \bibinfo {author} {\bibfnamefont {H.}~\bibnamefont {Zhang}}, \bibinfo
  {author} {\bibfnamefont {B.}~\bibnamefont {Yan}}, \bibinfo {author}
  {\bibfnamefont {X.-L.}\ \bibnamefont {Qi}}, \bibinfo {author} {\bibfnamefont
  {T.}~\bibnamefont {Frauenheim}}, \bibinfo {author} {\bibfnamefont
  {X.}~\bibnamefont {Dai}}, \bibinfo {author} {\bibfnamefont {Z.}~\bibnamefont
  {Fang}}, \ and\ \bibinfo {author} {\bibfnamefont {S.-C.}\ \bibnamefont
  {Zhang}},\ }\Doi {10.1103/PhysRevB.81.041307} {\bibfield  {journal} {\bibinfo
   {journal} {Physical Review B},\ }\textbf {\bibinfo {volume} {81}},\ \bibinfo
  {pages} {2} (\bibinfo {year} {2010})},\ ISSN \bibinfo {issn}
  {1098-0121}\BibitemShut {NoStop}%
\bibitem [{\citenamefont {Taskin}\ \emph {et~al.}(2012)\citenamefont {Taskin},
  \citenamefont {Sasaki}, \citenamefont {Segawa},\ and\ \citenamefont
  {Ando}}]{Taskin:2012}%
  \BibitemOpen
  \bibfield  {author} {\bibinfo {author} {\bibfnamefont {A.~A.}\ \bibnamefont
  {Taskin}}, \bibinfo {author} {\bibfnamefont {S.}~\bibnamefont {Sasaki}},
  \bibinfo {author} {\bibfnamefont {K.}~\bibnamefont {Segawa}}, \ and\ \bibinfo
  {author} {\bibfnamefont {Y.}~\bibnamefont {Ando}},\ }\Doi
  {10.1103/PhysRevLett.109.066803} {\bibfield  {journal} {\bibinfo  {journal}
  {Physical Review Letters},\ }\textbf {\bibinfo {volume} {109}},\ \bibinfo
  {pages} {066803} (\bibinfo {year} {2012})},\ ISSN \bibinfo {issn}
  {0031-9007}\BibitemShut {NoStop}%
\bibitem [{\citenamefont {Mahan}\ and\ \citenamefont
  {Sofo}(1996)}]{Mahan:1996p508}%
  \BibitemOpen
  \bibfield  {author} {\bibinfo {author} {\bibfnamefont {G.}~\bibnamefont
  {Mahan}}\ and\ \bibinfo {author} {\bibfnamefont {J.}~\bibnamefont {Sofo}},\
  }\href@noop {} {\bibfield  {journal} {\bibinfo  {journal} {Proceedings of the
  National Academy of Sciences},\ }\textbf {\bibinfo {volume} {93}},\ \bibinfo
  {pages} {7436} (\bibinfo {year} {1996})}\BibitemShut {NoStop}%
\bibitem [{\citenamefont {Zahn}\ \emph {et~al.}(1998)\citenamefont {Zahn},
  \citenamefont {Binder}, \citenamefont {Mertig}, \citenamefont {Zeller},\ and\
  \citenamefont {Dederichs}}]{Zahn:1998p15803}%
  \BibitemOpen
  \bibfield  {author} {\bibinfo {author} {\bibfnamefont {P.}~\bibnamefont
  {Zahn}}, \bibinfo {author} {\bibfnamefont {J.}~\bibnamefont {Binder}},
  \bibinfo {author} {\bibfnamefont {I.}~\bibnamefont {Mertig}}, \bibinfo
  {author} {\bibfnamefont {R.}~\bibnamefont {Zeller}}, \ and\ \bibinfo {author}
  {\bibfnamefont {P.}~\bibnamefont {Dederichs}},\ }\Doi
  {10.1103/PhysRevLett.80.4309} {\bibfield  {journal} {\bibinfo  {journal}
  {Phys. Rev. Lett.},\ }\textbf {\bibinfo {volume} {80}},\ \bibinfo {pages}
  {4309} (\bibinfo {year} {1998})}\BibitemShut {NoStop}%
\bibitem [{Note1()}]{Note1}%
  \BibitemOpen
  \bibinfo {note} {The fixed charge $N$ at varying $\mu (T)$ is determined by
  an integration over the density of states $g(E)$ $N=\DOTSI \intop \ilimits@
  \limits _{\mu -\Delta E}^{\protect \text {VB}^{max}} \protect \text {d}E
  \protect \tmspace +\thinmuskip {.1667em} g(E) [f(\mu ,T)-1]+ \DOTSI \intop
  \ilimits@ \limits _{\protect \text {CB}^{min}}^{\mu +\Delta E} \protect \text
  {d}E \protect \tmspace +\thinmuskip {.1667em} g(E) f(\mu ,T)$}\BibitemShut
  {NoStop}%
\bibitem [{\citenamefont {Zahn}\ \emph {et~al.}(2011)\citenamefont {Zahn},
  \citenamefont {Hinsche}, \citenamefont {Yavorsky},\ and\ \citenamefont
  {Mertig}}]{Zahn:2011p15523}%
  \BibitemOpen
  \bibfield  {author} {\bibinfo {author} {\bibfnamefont {P.}~\bibnamefont
  {Zahn}}, \bibinfo {author} {\bibfnamefont {N.}~\bibnamefont {Hinsche}},
  \bibinfo {author} {\bibfnamefont {B.}~\bibnamefont {Yavorsky}}, \ and\
  \bibinfo {author} {\bibfnamefont {I.}~\bibnamefont {Mertig}},\ }\Doi
  {10.1088/0953-8984/23/50/505504} {\bibfield  {journal} {\bibinfo  {journal}
  {J. Phys.: Condens. Matter},\ }\textbf {\bibinfo {volume} {23}},\ \bibinfo
  {pages} {505504} (\bibinfo {year} {2011})}\BibitemShut {NoStop}%
\bibitem [{\citenamefont {Yavorsky}\ \emph {et~al.}(2011)\citenamefont
  {Yavorsky}, \citenamefont {Hinsche}, \citenamefont {Mertig},\ and\
  \citenamefont {Zahn}}]{Yavorsky:2011p15466}%
  \BibitemOpen
  \bibfield  {author} {\bibinfo {author} {\bibfnamefont {B.~Y.}\ \bibnamefont
  {Yavorsky}}, \bibinfo {author} {\bibfnamefont {N.}~\bibnamefont {Hinsche}},
  \bibinfo {author} {\bibfnamefont {I.}~\bibnamefont {Mertig}}, \ and\ \bibinfo
  {author} {\bibfnamefont {P.}~\bibnamefont {Zahn}},\ }\Doi
  {10.1103/PhysRevB.84.165208} {\bibfield  {journal} {\bibinfo  {journal}
  {Physical Review B},\ }\textbf {\bibinfo {volume} {84}},\ \bibinfo {pages}
  {165208} (\bibinfo {year} {2011})}\BibitemShut {NoStop}%
\bibitem [{Note2()}]{Note2}%
  \BibitemOpen
  \bibinfo {note} {Throughout this study a threshold of 0.90 was
  used.}\BibitemShut {Stop}%
\bibitem [{\citenamefont {Henk}\ \emph
  {et~al.}(2012){\natexlab{b}}\citenamefont {Henk}, \citenamefont {Ernst},
  \citenamefont {Eremeev}, \citenamefont {Chulkov}, \citenamefont
  {Maznichenko},\ and\ \citenamefont {Mertig}}]{Henk:2012}%
  \BibitemOpen
  \bibfield  {author} {\bibinfo {author} {\bibfnamefont {J.}~\bibnamefont
  {Henk}}, \bibinfo {author} {\bibfnamefont {A.}~\bibnamefont {Ernst}},
  \bibinfo {author} {\bibfnamefont {S.}~\bibnamefont {Eremeev}}, \bibinfo
  {author} {\bibfnamefont {E.}~\bibnamefont {Chulkov}}, \bibinfo {author}
  {\bibfnamefont {I.}~\bibnamefont {Maznichenko}}, \ and\ \bibinfo {author}
  {\bibfnamefont {I.}~\bibnamefont {Mertig}},\ }\Doi
  {10.1103/PhysRevLett.108.206801} {\bibfield  {journal} {\bibinfo  {journal}
  {Physical Review Letters},\ }\textbf {\bibinfo {volume} {108}} (\bibinfo
  {year} {2012}{\natexlab{b}})},\ ISSN \bibinfo {issn} {0031-9007},\ \doi
  {10.1103/PhysRevLett.108.206801}\BibitemShut {NoStop}%
\bibitem [{\citenamefont {Alpichshev}\ \emph {et~al.}(2010)\citenamefont
  {Alpichshev}, \citenamefont {Analytis}, \citenamefont {Chu}, \citenamefont
  {Fisher}, \citenamefont {Chen}, \citenamefont {Shen}, \citenamefont {Fang},\
  and\ \citenamefont {Kapitulnik}}]{Alpichshev:2010p15262}%
  \BibitemOpen
  \bibfield  {author} {\bibinfo {author} {\bibfnamefont {Z.}~\bibnamefont
  {Alpichshev}}, \bibinfo {author} {\bibfnamefont {J.~G.}\ \bibnamefont
  {Analytis}}, \bibinfo {author} {\bibfnamefont {J.-H.}\ \bibnamefont {Chu}},
  \bibinfo {author} {\bibfnamefont {I.~R.}\ \bibnamefont {Fisher}}, \bibinfo
  {author} {\bibfnamefont {Y.~L.}\ \bibnamefont {Chen}}, \bibinfo {author}
  {\bibfnamefont {Z.~X.}\ \bibnamefont {Shen}}, \bibinfo {author}
  {\bibfnamefont {A.}~\bibnamefont {Fang}}, \ and\ \bibinfo {author}
  {\bibfnamefont {A.}~\bibnamefont {Kapitulnik}},\ }\Doi
  {10.1103/PhysRevLett.104.016401} {\bibfield  {journal} {\bibinfo  {journal}
  {Phys. Rev. Lett.},\ }\textbf {\bibinfo {volume} {104}},\ \bibinfo {pages}
  {016401} (\bibinfo {year} {2010})}\BibitemShut {NoStop}%
\bibitem [{\citenamefont {Taskin}\ \emph
  {et~al.}(2011){\natexlab{b}}\citenamefont {Taskin}, \citenamefont {Ren},
  \citenamefont {Sasaki}, \citenamefont {Segawa},\ and\ \citenamefont
  {Ando}}]{Taskin:2011}%
  \BibitemOpen
  \bibfield  {author} {\bibinfo {author} {\bibfnamefont {A.~A.}\ \bibnamefont
  {Taskin}}, \bibinfo {author} {\bibfnamefont {Z.}~\bibnamefont {Ren}},
  \bibinfo {author} {\bibfnamefont {S.}~\bibnamefont {Sasaki}}, \bibinfo
  {author} {\bibfnamefont {K.}~\bibnamefont {Segawa}}, \ and\ \bibinfo {author}
  {\bibfnamefont {Y.}~\bibnamefont {Ando}},\ }\Doi
  {10.1103/PhysRevLett.107.016801} {\bibfield  {journal} {\bibinfo  {journal}
  {Physical Review Letters},\ }\textbf {\bibinfo {volume} {107}},\ \bibinfo
  {pages} {016801} (\bibinfo {year} {2011}{\natexlab{b}})},\ ISSN \bibinfo
  {issn} {0031-9007}\BibitemShut {NoStop}%
\bibitem [{\citenamefont {Hashibon}\ and\ \citenamefont
  {Els\"{a}sser}(2011)}]{Hashibon:2011}%
  \BibitemOpen
  \bibfield  {author} {\bibinfo {author} {\bibfnamefont {A.}~\bibnamefont
  {Hashibon}}\ and\ \bibinfo {author} {\bibfnamefont {C.}~\bibnamefont
  {Els\"{a}sser}},\ }\Doi {10.1103/PhysRevB.84.144117} {\bibfield  {journal}
  {\bibinfo  {journal} {Physical Review B},\ }\textbf {\bibinfo {volume}
  {84}},\ \bibinfo {pages} {144117} (\bibinfo {year} {2011})},\ ISSN \bibinfo
  {issn} {1098-0121}\BibitemShut {NoStop}%
\bibitem [{Note3()}]{Note3}%
  \BibitemOpen
  \bibinfo {note} {We found additional small contributions to the surface state
  at energies $E \le E_{\protect \text {D}}-\protect \unit [250]{meV}$. The
  latter, do not noticeably contribute to the thermoelectric transport even at
  large hole charge carrier concentrations and high temperatures and can be
  omitted.}\BibitemShut {Stop}%
\bibitem [{\citenamefont {Peranio}\ \emph {et~al.}(2006)\citenamefont
  {Peranio}, \citenamefont {Eibl},\ and\ \citenamefont {Nurnus}}]{Peranio2006}%
  \BibitemOpen
  \bibfield  {author} {\bibinfo {author} {\bibfnamefont {N.}~\bibnamefont
  {Peranio}}, \bibinfo {author} {\bibfnamefont {O.}~\bibnamefont {Eibl}}, \
  and\ \bibinfo {author} {\bibfnamefont {J.}~\bibnamefont {Nurnus}},\ }\Doi
  {10.1063/1.2375016} {\bibfield  {journal} {\bibinfo  {journal} {Journal of
  Applied Physics},\ }\textbf {\bibinfo {volume} {100}},\ \bibinfo {pages}
  {114306} (\bibinfo {year} {2006})}\BibitemShut {NoStop}%
\bibitem [{\citenamefont {Zastrow}\ \emph {et~al.}(2013)\citenamefont
  {Zastrow}, \citenamefont {Gooth}, \citenamefont {Boehnert}, \citenamefont
  {Heiderich}, \citenamefont {Toellner}, \citenamefont {Heimann}, \citenamefont
  {Schulz},\ and\ \citenamefont {Nielsch}}]{Zastrow:2013p16357}%
  \BibitemOpen
  \bibfield  {author} {\bibinfo {author} {\bibfnamefont {S.}~\bibnamefont
  {Zastrow}}, \bibinfo {author} {\bibfnamefont {J.}~\bibnamefont {Gooth}},
  \bibinfo {author} {\bibfnamefont {T.}~\bibnamefont {Boehnert}}, \bibinfo
  {author} {\bibfnamefont {S.}~\bibnamefont {Heiderich}}, \bibinfo {author}
  {\bibfnamefont {W.}~\bibnamefont {Toellner}}, \bibinfo {author}
  {\bibfnamefont {S.}~\bibnamefont {Heimann}}, \bibinfo {author} {\bibfnamefont
  {S.}~\bibnamefont {Schulz}}, \ and\ \bibinfo {author} {\bibfnamefont
  {K.}~\bibnamefont {Nielsch}},\ }\href@noop {} {\bibfield  {journal} {\bibinfo
   {journal} {Semiconductor Science and Technology},\ }\textbf {\bibinfo
  {volume} {28}},\ \bibinfo {pages} {035010} (\bibinfo {year}
  {2013})}\BibitemShut {NoStop}%
\bibitem [{\citenamefont {Boulouz}\ \emph {et~al.}(2001)\citenamefont
  {Boulouz}, \citenamefont {Chakraborty}, \citenamefont {Giani}, \citenamefont
  {Delannoy}, \citenamefont {Boyer},\ and\ \citenamefont
  {Schumann}}]{Boulouz:2001p16385}%
  \BibitemOpen
  \bibfield  {author} {\bibinfo {author} {\bibfnamefont {A.}~\bibnamefont
  {Boulouz}}, \bibinfo {author} {\bibfnamefont {S.}~\bibnamefont
  {Chakraborty}}, \bibinfo {author} {\bibfnamefont {A.}~\bibnamefont {Giani}},
  \bibinfo {author} {\bibfnamefont {F.}~\bibnamefont {Delannoy}}, \bibinfo
  {author} {\bibfnamefont {A.}~\bibnamefont {Boyer}}, \ and\ \bibinfo {author}
  {\bibfnamefont {J.}~\bibnamefont {Schumann}},\ }\href@noop {} {\bibfield
  {journal} {\bibinfo  {journal} {Journal of Applied Physics},\ }\textbf
  {\bibinfo {volume} {89}},\ \bibinfo {pages} {5009} (\bibinfo {year}
  {2001})}\BibitemShut {NoStop}%
\bibitem [{\citenamefont {Lee}(2009)}]{Lee:2009p16355}%
  \BibitemOpen
  \bibfield  {author} {\bibinfo {author} {\bibfnamefont {D.-H.}\ \bibnamefont
  {Lee}},\ }\href@noop {} {\bibfield  {journal} {\bibinfo  {journal} {Phys.
  Rev. Lett.},\ }\textbf {\bibinfo {volume} {103}},\ \bibinfo {pages} {196804}
  (\bibinfo {year} {2009})}\BibitemShut {NoStop}%
\bibitem [{Note4()}]{Note4}%
  \BibitemOpen
  \bibinfo {note} {Unfortunately, the precise numerical determination of the
  electron-phonon coupling on an \protect \textit {ab initio} level is quite
  demanding for semi-conducting materials. The \protect \textsc
  {\`{E}liashberg} function has to be determined for a rather large numbers of
  chemical potentials.}\BibitemShut {Stop}%
\bibitem [{\citenamefont {Scheidemantel}\ \emph {et~al.}(2003)\citenamefont
  {Scheidemantel}, \citenamefont {Ambrosch-Draxl}, \citenamefont {Thonhauser},
  \citenamefont {Badding},\ and\ \citenamefont
  {Sofo}}]{Scheidemantel:2003p14961}%
  \BibitemOpen
  \bibfield  {author} {\bibinfo {author} {\bibfnamefont {T.}~\bibnamefont
  {Scheidemantel}}, \bibinfo {author} {\bibfnamefont {C.}~\bibnamefont
  {Ambrosch-Draxl}}, \bibinfo {author} {\bibfnamefont {T.}~\bibnamefont
  {Thonhauser}}, \bibinfo {author} {\bibfnamefont {J.}~\bibnamefont {Badding}},
  \ and\ \bibinfo {author} {\bibfnamefont {J.}~\bibnamefont {Sofo}},\
  }\href@noop {} {\bibfield  {journal} {\bibinfo  {journal} {Physical Review
  B},\ }\textbf {\bibinfo {volume} {68}},\ \bibinfo {pages} {125210} (\bibinfo
  {year} {2003})}\BibitemShut {NoStop}%
\bibitem [{\citenamefont {Stordeur}\ \emph {et~al.}(1988)\citenamefont
  {Stordeur}, \citenamefont {St{\"o}lzer}, \citenamefont {Sobotta},\ and\
  \citenamefont {Riede}}]{Stordeur:1988p15503}%
  \BibitemOpen
  \bibfield  {author} {\bibinfo {author} {\bibfnamefont {M.}~\bibnamefont
  {Stordeur}}, \bibinfo {author} {\bibfnamefont {M.}~\bibnamefont
  {St{\"o}lzer}}, \bibinfo {author} {\bibfnamefont {H.}~\bibnamefont
  {Sobotta}}, \ and\ \bibinfo {author} {\bibfnamefont {V.}~\bibnamefont
  {Riede}},\ }\Doi {10.1002/pssb.2221500120} {\bibfield  {journal} {\bibinfo
  {journal} {physica status solidi (b)},\ }\textbf {\bibinfo {volume} {150}},\
  \bibinfo {pages} {165} (\bibinfo {year} {1988})}\BibitemShut {NoStop}%
\bibitem [{\citenamefont {Mahan}(1990)}]{Mahan:1990p16105}%
  \BibitemOpen
  \bibfield  {author} {\bibinfo {author} {\bibfnamefont {G.~D.}\ \bibnamefont
  {Mahan}},\ }\href@noop {} {\emph {\bibinfo {title} {{Many particle
  physics}}}}\ (\bibinfo  {publisher} {Plenum Press},\ \bibinfo {address} {New
  York},\ \bibinfo {year} {1990})\BibitemShut {NoStop}%
\bibitem [{\citenamefont {Fabian}\ and\ \citenamefont
  {Sarma}(1999)}]{Fabian:1999p16116}%
  \BibitemOpen
  \bibfield  {author} {\bibinfo {author} {\bibfnamefont {J.}~\bibnamefont
  {Fabian}}\ and\ \bibinfo {author} {\bibfnamefont {S.~D.}\ \bibnamefont
  {Sarma}},\ }\href@noop {} {\bibfield  {journal} {\bibinfo  {journal}
  {Physical Review Letters},\ }\textbf {\bibinfo {volume} {83}},\ \bibinfo
  {pages} {1211} (\bibinfo {year} {1999})}\BibitemShut {NoStop}%
\bibitem [{\citenamefont {Shoemake}\ \emph {et~al.}(1969)\citenamefont
  {Shoemake}, \citenamefont {Rayne},\ and\ \citenamefont
  {Ure}}]{Shoemake:1969p16343}%
  \BibitemOpen
  \bibfield  {author} {\bibinfo {author} {\bibfnamefont {G.}~\bibnamefont
  {Shoemake}}, \bibinfo {author} {\bibfnamefont {J.}~\bibnamefont {Rayne}}, \
  and\ \bibinfo {author} {\bibfnamefont {R.}~\bibnamefont {Ure}},\ }\href@noop
  {} {\bibfield  {journal} {\bibinfo  {journal} {Physical Review},\ }\textbf
  {\bibinfo {volume} {185}},\ \bibinfo {pages} {1046} (\bibinfo {year}
  {1969})}\BibitemShut {NoStop}%
\bibitem [{\citenamefont {McMillan}(1968)}]{McMillan:1968p16342}%
  \BibitemOpen
  \bibfield  {author} {\bibinfo {author} {\bibfnamefont {W.}~\bibnamefont
  {McMillan}},\ }\href@noop {} {\bibfield  {journal} {\bibinfo  {journal}
  {Physical Review},\ }\textbf {\bibinfo {volume} {167}},\ \bibinfo {pages}
  {331} (\bibinfo {year} {1968})}\BibitemShut {NoStop}%
\bibitem [{\citenamefont {Chen}\ \emph {et~al.}(2013)\citenamefont {Chen},
  \citenamefont {Xie}, \citenamefont {Feng}, \citenamefont {Yi}, \citenamefont
  {Liang}, \citenamefont {He}, \citenamefont {Mou}, \citenamefont {He},
  \citenamefont {Peng}, \citenamefont {Liu}, \citenamefont {Liu}, \citenamefont
  {Zhao}, \citenamefont {Liu}, \citenamefont {Dong}, \citenamefont {Zhang},
  \citenamefont {Yu}, \citenamefont {Wang}, \citenamefont {Peng}, \citenamefont
  {Wang}, \citenamefont {Zhang}, \citenamefont {Yang}, \citenamefont {Chen},
  \citenamefont {Xu},\ and\ \citenamefont {Zhou}}]{Chen:2013p16344}%
  \BibitemOpen
  \bibfield  {author} {\bibinfo {author} {\bibfnamefont {C.}~\bibnamefont
  {Chen}}, \bibinfo {author} {\bibfnamefont {Z.}~\bibnamefont {Xie}}, \bibinfo
  {author} {\bibfnamefont {Y.}~\bibnamefont {Feng}}, \bibinfo {author}
  {\bibfnamefont {H.}~\bibnamefont {Yi}}, \bibinfo {author} {\bibfnamefont
  {A.}~\bibnamefont {Liang}}, \bibinfo {author} {\bibfnamefont
  {S.}~\bibnamefont {He}}, \bibinfo {author} {\bibfnamefont {D.}~\bibnamefont
  {Mou}}, \bibinfo {author} {\bibfnamefont {J.}~\bibnamefont {He}}, \bibinfo
  {author} {\bibfnamefont {Y.}~\bibnamefont {Peng}}, \bibinfo {author}
  {\bibfnamefont {X.}~\bibnamefont {Liu}}, \bibinfo {author} {\bibfnamefont
  {Y.}~\bibnamefont {Liu}}, \bibinfo {author} {\bibfnamefont {L.}~\bibnamefont
  {Zhao}}, \bibinfo {author} {\bibfnamefont {G.}~\bibnamefont {Liu}}, \bibinfo
  {author} {\bibfnamefont {X.}~\bibnamefont {Dong}}, \bibinfo {author}
  {\bibfnamefont {J.}~\bibnamefont {Zhang}}, \bibinfo {author} {\bibfnamefont
  {L.}~\bibnamefont {Yu}}, \bibinfo {author} {\bibfnamefont {X.}~\bibnamefont
  {Wang}}, \bibinfo {author} {\bibfnamefont {Q.}~\bibnamefont {Peng}}, \bibinfo
  {author} {\bibfnamefont {Z.}~\bibnamefont {Wang}}, \bibinfo {author}
  {\bibfnamefont {S.}~\bibnamefont {Zhang}}, \bibinfo {author} {\bibfnamefont
  {F.}~\bibnamefont {Yang}}, \bibinfo {author} {\bibfnamefont {C.}~\bibnamefont
  {Chen}}, \bibinfo {author} {\bibfnamefont {Z.}~\bibnamefont {Xu}}, \ and\
  \bibinfo {author} {\bibfnamefont {X.~J.}\ \bibnamefont {Zhou}},\ }\href@noop
  {} {\bibfield  {journal} {\bibinfo  {journal} {Scientific Reports},\ }\textbf
  {\bibinfo {volume} {3}} (\bibinfo {year} {2013})}\BibitemShut {NoStop}%
\bibitem [{\citenamefont {Huang}(2012)}]{Huang:2012p16206}%
  \BibitemOpen
  \bibfield  {author} {\bibinfo {author} {\bibfnamefont {G.~Q.}\ \bibnamefont
  {Huang}},\ }\href@noop {} {\bibfield  {journal} {\bibinfo  {journal} {EPL
  (Europhysics Letters)},\ }\textbf {\bibinfo {volume} {100}},\ \bibinfo
  {pages} {17001} (\bibinfo {year} {2012})}\BibitemShut {NoStop}%
\bibitem [{\citenamefont {Giraud}\ \emph {et~al.}(2012)\citenamefont {Giraud},
  \citenamefont {Kundu},\ and\ \citenamefont {Egger}}]{Giraud:2012p16101}%
  \BibitemOpen
  \bibfield  {author} {\bibinfo {author} {\bibfnamefont {S.}~\bibnamefont
  {Giraud}}, \bibinfo {author} {\bibfnamefont {A.}~\bibnamefont {Kundu}}, \
  and\ \bibinfo {author} {\bibfnamefont {R.}~\bibnamefont {Egger}},\
  }\href@noop {} {\bibfield  {journal} {\bibinfo  {journal} {Physical Review
  B},\ }\textbf {\bibinfo {volume} {85}},\ \bibinfo {pages} {035441} (\bibinfo
  {year} {2012})}\BibitemShut {NoStop}%
\bibitem [{\citenamefont {Pan}\ \emph {et~al.}(2012)\citenamefont {Pan},
  \citenamefont {Fedorov}, \citenamefont {Gardner}, \citenamefont {Lee},
  \citenamefont {Chu},\ and\ \citenamefont {Valla}}]{Pan:2012p16341}%
  \BibitemOpen
  \bibfield  {author} {\bibinfo {author} {\bibfnamefont {Z.-H.}\ \bibnamefont
  {Pan}}, \bibinfo {author} {\bibfnamefont {A.~V.}\ \bibnamefont {Fedorov}},
  \bibinfo {author} {\bibfnamefont {D.}~\bibnamefont {Gardner}}, \bibinfo
  {author} {\bibfnamefont {Y.~S.}\ \bibnamefont {Lee}}, \bibinfo {author}
  {\bibfnamefont {S.}~\bibnamefont {Chu}}, \ and\ \bibinfo {author}
  {\bibfnamefont {T.}~\bibnamefont {Valla}},\ }\href@noop {} {\bibfield
  {journal} {\bibinfo  {journal} {Phys. Rev. Lett.},\ }\textbf {\bibinfo
  {volume} {108}},\ \bibinfo {pages} {187001} (\bibinfo {year}
  {2012})}\BibitemShut {NoStop}%
\end{thebibliography}%

\end{document}